\documentclass[12pt]{article}
\usepackage{amsmath,amssymb,eucal}
\font\tenrm=cmr10

\font\bss=cmssdc10 at 12 pt
\font\bigss=cmssdc10 scaled 2300

\font\cmsslll=cmss10 at 14 pt

\renewcommand{\a}{\alpha}
\renewcommand{\b}{\beta}

\newcommand{\e}{\epsilon}
\newcommand{\f}{\varphi}
\newcommand{\g}{\gamma}

\renewcommand{\i}{\iota}

\renewcommand{\o}{\omega}

\newcommand{\s}{\sigma}
\renewcommand{\t}{\tau}

\newcommand{\G}{\Gamma}

\newcommand{\bC}{\mathbb{C}}
\newcommand{\bR}{\mathbb{R}}
\newcommand{\bZ}{\mathbb{Z}}
\newcommand{\bH}{\mathbb{H}}
\newcommand{\bN}{\mathbb{N}}

\newcommand{\bS}{\mathbb{S}}
\newcommand{\bV}{\mathbb{V}}
\newcommand{\bW}{\mathbb{W}}
\newcommand{\bK}{\mathbb{K}}

\newcommand{\ga}{\mathfrak{a}}

\renewcommand{\gg}{\mathfrak{g}}
\newcommand{\gh}{\mathfrak{h}}
\newcommand{\gk}{\mathfrak{k}}
\newcommand{\gl}{\mathfrak{l}}

\newcommand{\gS}{\mathfrak{S}}

\newcommand{\so}{\mathfrak{so}}
\newcommand{\su}{\mathfrak{su}}

\newcommand{\gsp}{\mathfrak{sp}}

\newcommand{\conf}{\mathfrak{conf}}
\newcommand{\osp}{\mathfrak{osp}}
\newcommand{\spo}{\mathfrak{spo}}

\newcommand\Spin{\mathrm{Spin}}

\newcommand{\cc}{\mathcal{C}}

\newcommand{\Cl}{\mathcal{C}l}

\newcommand\fr[2]{\tfrac{#1}{#2}}

\renewcommand{\square}{\kern1pt\vbox
               {\hrule height 0.6pt\hbox{\vrule width 0.6pt\hskip 3pt
    \vbox{\vskip 6pt}\hskip 3pt\vrule width 0.6pt}\hrule height0.6pt}
    \kern1pt}

\newcommand{\La}{\wedge}
\newcommand{\ra}{\rightarrow}

\DeclareMathOperator\End{End\;}
\DeclareMathOperator\vol{vol\;}

\DeclareMathOperator\ad{ad\;}
\DeclareMathOperator\Bil{Bil}

\newcommand{\wt}{\widetilde}
\newcommand{\wh}{\widehat}
\newcommand{\ol}{\overline}

\newtheorem{Th}{Theorem}
\newtheorem{Prop}{Proposition}
\newtheorem{Cor}{Corollary}
\newtheorem{Lem}{Lemma}
\newtheorem{Def}{Definition}

\newcommand{\bt}{\begin{Th}\ \ }
\newcommand{\et}{\end{Th}}
\newcommand{\bp}{\begin{Prop}\ \ }
\newcommand{\ep}{\end{Prop}}
\newcommand{\bc}{\begin{Cor}\ \ }
\newcommand{\ec}{\end{Cor}}
\newcommand{\bl}{\begin{Lem}\ \ }
\newcommand{\el}{\end{Lem}}
\newcommand{\bd}{\begin{Def}\ \ }
\newcommand{\ed}{\end{Def}}

\newcommand{\pf}{\noindent{\it Proof:\ \ }}
\newcommand{\qed}{\hfill\square}

\newcommand{\op}{\oplus}
\newcommand{\ot}{\otimes}

\newcommand{\bbS}{\overline{\mathbb S}}

\newcommand\re[1]{(\ref{#1})}
\newcommand{\arr}{\begin{array}{rlll}}
\newcommand{\ea}{\end{array}}
\newcommand{\bea}{\begin{eqnarray}}
\newcommand{\eea}{\end{eqnarray}}
\newcommand{\bean}{\begin{eqnarray*}}
\newcommand{\eean}{\end{eqnarray*}}

\catcode`@=11
\@addtoreset{equation}{section}
\catcode`@=12
\newcommand{\ft}[2]{{\textstyle\frac{#1}{#2}}}
\def\rmi{{\rm i}}

\newsavebox{\uuunit}
\sbox{\uuunit}
    {\setlength{\unitlength}{0.825em}
     \begin{picture}(0.6,0.7)
        \thinlines
        \put(0,0){\line(1,0){0.5}}
        \put(0.15,0){\line(0,1){0.7}}
        \put(0.35,0){\line(0,1){0.8}}
       \multiput(0.3,0.8)(-0.04,-0.02){12}{\rule{0.5pt}{0.5pt}}
     \end {picture}}
\newcommand {\unity}{\mathord{\!\usebox{\uuunit}}}
\begin{document}
\begin{titlepage}
\rightline{hep-th/0311107}
\rightline{KUL-TF-03/16}
\vskip 1.5 true cm
\begin{center}
{\bigss  Polyvector Super-Poincar{\'e} Algebras}
\vskip 1.0 true cm
{\cmsslll   Dmitri V.\ Alekseevsky}\\[3pt]
{\tenrm   Dept. of Mathematics, University of Hull,
Cottingham Road, Hull, HU6 7RX, UK \\
D.V.Alekseevsky@maths.hull.ac.uk}
\vskip 0.2 true cm
{\cmsslll    Vicente Cort{\'e}s} \\[3pt]
{\tenrm    Institut \'Elie Cartan de Math{\'e}matiques,
Universit{\'e} Henri Poincar{\'e} - Nancy 1,
B.P. 239,\\ F-54506 Vandoeuvre-l{\`e}s-Nancy Cedex, France\\
cortes@iecn.u-nancy.fr}
\vskip 0.2 true cm
{\cmsslll    Chandrashekar Devchand}\\[3pt]
{\tenrm   Mathematisches Institut der Universit{\"a}t Bonn,
Beringstra\ss e 1, D-53115 Bonn, Germany\\
devchand@math.uni-bonn.de}
\vskip 0.2 true cm
{\cmsslll   Antoine Van Proeyen}  \\[3pt]
{\tenrm   Instituut voor Theoretische Fysica, Katholieke Universiteit
Leuven, Celestijnenlaan 200D,\\ B-30001 Leuven, Belgium \\
Antoine.VanProeyen@fys.kuleuven.ac.be}
\end{center}
\vskip 1.0 true cm
%%%%%%%%%%%%%%%%%%%%%%%%%%%%%%%%%%%%%%%%%%%%%%%%%%%%%%%%
\begin{abstract}
\noindent A class of $\bZ_2$-graded  Lie algebra and Lie superalgebra
extensions of the pseudo-orthogonal algebra of a spacetime of arbitrary
dimension and signature is investigated. They have the form $\gg = \gg_0
+ \gg_1$, with $\gg_0 = \so(V) + W_0$ and $\gg_1=W_1$, where the algebra
of generalized translations $W = W_0 + W_1$ is the maximal solvable ideal
of $\gg$, $W_0$ is generated by $W_1$ and commutes with $W$. Choosing
$W_1$ to be a spinorial $\so(V)$-module (a sum of an arbitrary number of
spinors and semispinors), we prove that $W_0$ consists of polyvectors,
i.e.\ all the irreducible $\so(V)$-submodules of $W_0$ are submodules of
$\La V$. We provide a classification of such  Lie (super)algebras for all
dimensions and signatures. The problem reduces to the classification of
$\so(V)$-invariant $\La^kV$-valued bilinear forms on the spinor module
$S$.
\end{abstract}

\end{titlepage}
{\small \tableofcontents}

\section{Introduction}

A {\it superextension} of  a Lie algebra $\gh$ is a Lie superalgebra
$\gg =\gg_0 + \gg_1\;$,  such that $\gh \subset \gg_0$. If the Lie algebra
$\gg_0 \supset \gh$ and a $\gg_0$-module $\gg_1$ are given, then a
superextension is determined by the Lie superbracket in the odd part, which
is  a $\gg_0$-equivariant linear map
\begin{equation}
  \vee^2 \gg_1 \ra \gg_0\  ,
\end{equation} satisfying the Jacobi identity for $X,Y,Z \in \gg_1$, where $\vee$
denotes the symmetric tensor product. Similarly, a {\it $\bZ_2$-graded
extension} (or simply {\it Lie extension}) of $\gh$ is a $\bZ_2$-graded
Lie algebra $\gg$, i.e.\ a Lie algebra with a $\bZ_2$-grading $\gg =\gg_0
+ \gg_1$ compatible with the Lie bracket: $ \left[ \gg_\a , \gg_\b
\right]\subset \gg_{\a+\b}\,,\,\a,\b\in\bZ/2\bZ\,$,  such that $\gg_0
\supset \gh$. As above, a $\bZ_2$-graded extension is determined by the
Lie bracket in $\,\gg_1$, which defines a $\gg_0$-equivariant linear map,
\begin{equation}
 \La^2 \gg_1 \ra \gg_0\ ,
\end{equation}
satisfying the Jacobi identity. For instance, consider a super vector space
$V_0 + V_1$ endowed with a scalar superproduct $g = g_0 + g_1$, i.e.\
$g_0$ is a (possibly indefinite) scalar product on $V_0$ and $g_1$ is a
nondegenerate skewsymmetric bilinear form on $V_1$. The Lie algebra
$\gh = \gg_0 = \so(V_0) \oplus \gsp(V_1)$ of infinitesimal even automorphisms
of  $(V_0+V_1,g)$ has a natural extension with $\gg_1 = V_0 \otimes V_1$,
where the Lie superbracket is given by:
\[ [ v_0\otimes v_1 , v_0'\otimes v_1'] := g_1(v_1,v_1')v_0\wedge v_0'
+ g_0(v_0,v_0') v_1 \vee v_1'\, .\]
This is the orthosymplectic Lie superalgebra $\osp(V_0|V_1)$. One can also
define an analogous Lie superalgebra  $\spo(V_0|V_1)$, starting from a
symplectic super vector space   $(V_0+V_1, \o = \o_0 + \o_1)$, such that
$\spo(V_0|V_1)=\osp(V_1|V_0)$.

Similarly, for a $\bZ_2$-graded vector space $V_0 + V_1$ endowed with
a scalar product $g = g_0 + g_1$ (respectively, a
symplectic form $\o = \o_0 + \o_1$) we have a natural $\bZ_2$-graded
extension $\gg = \gg_0 + \gg_1 =
\so(V_0+V_1)$ (respectively,  $\gg = \gsp(V_0+V_1)$)
of the Lie algebra $\gh = \gg_0 = \so(V_0) \oplus \so(V_1)$ (respectively,
of $\gh =\gsp(V_0) \oplus \gsp(V_1)$).

For a pseudo-Euclidean space-time $V =\bR^{p,q}$ (with $p$ positive
and $q$ negative directions), Nahm \cite{N} classified
superextensions $\gg$ of  the
pseudo-orthogonal Lie algebra $\so(V)$ under the assumptions that
$q \le 2$, $\gg$ is simple,  $\gg_0$ is a direct sum of ideals,
$\gg_0= \so(V) \op \gk\,$, where $\gk$ is reductive and
$\gg_1$ is a spinorial module (i.e.\  its irreducible
summands are spinors or semi-spinors). These algebras for $q=2$ are usually
considered as superconformal algebras for Minkowski spacetimes,
in virtue of the identification $\conf(p-1,1) = \so(p,2)$.

In this paper, we shall consider both super and Lie extensions (which
we call\linebreak[3] {\it $\e$-extensions}) of the pseudo-orthogonal
Lie algebra
$\so(V)$, with $\e{=}+1$ corresponding to
superextensions and  $\e{=}-1$  to Lie extensions. Here $V=\bR^{p,q}$ or
$V = \bC^n$ is a vector space endowed with a scalar product. In the case
$\gg_0 =  \so(V) + V$ (Poincar{\'e} Lie algebra),
$\e$-extensions  $\gg = \gg_0 + \gg_1$ such that $\gg_1$ is a spinorial module
and $[\gg_1,\gg_1] \subset V$  were classified in \cite{AC}.
In the case $\e = -1$ such extensions clearly do not respect the
conventional  field theoretical  spin--statistics relationship.
However, in order to classify super-Poincar{\'e} algebras ($\e=+1$) with an
arbitrary number of irreducible spinorial submodules in $\gg_1$
we need to classify Lie extensions  as well as superextensions with
irreducible $\gg_1$.

We study  $\bZ_2$-graded  Lie algebras and Lie superalgebras,
$\,\gg = \gg_0 + \gg_1$ where $\gg_0 =\so(V) + W_0\ ,\  \gg_1=W_1$,
such that $\so(V)$ is a maximal semisimple Lie subalgebra of $\gg$ and
$W = W_0 + W_1$ is its maximal solvable ideal. If  $W_0$ contains
$[W_1,W_1]$ and commutes with $W$, we  call  $\gg$ an $\e$-extension
of translational type. If moreover,
$W_0 = [W_1,W_1]$, we call $\gg$ an $\e$-transalgebra.
Our main result is the classification
of {\it $\e$-extended polyvector Poincar{\'e} algebras}, i.e.\
$\e$-extensions of translational type in the case when  $W_1=S$,
the spinor  $\so(V)$-module, or, more generally, an arbitrary spinorial
module.   Here $V$ is  an arbitrary pseudo-Euclidean vector space
$\bR^{p,q}$. We prove  that, under these assumptions, any irreducible
$\so(V)$-submodule of $W_0$ is of the form $\La^kV$ or $\La^m_\pm V$,
where $m = (p+q)/2$ and   $\La^m_\pm V$ are the eigenspaces of the Hodge star
operator on  $\La^mV$.

If $\gg= \so(V) + W_0 + S$ is an $\e$-transalgebra, then the (super)
Lie bracket defines an $\so(V)$-equivariant surjective map
$\G_{W_0}: S\ot S\ra W_0$.
If $K$ is the kernel of this map, then  there exists a complementary
submodule $\wt{W}_0$ such that $S\ot S= \wt{W}_0+K$ and we can
identify $\wt{W}_0$ with $W_0$. We note that we can choose
$\wt{W}_0 \subset S\wedge S$ in the
Lie algebra case and $\wt{W}_0 \subset S\vee S$ in the Lie superalgebra case.
Conversely, if we have a decomposition $S\ot S = W_0+K$ into a
sum of two $\so(V)$-submodules and moreover $W_0 \subset S\wedge S$
or $W_0 \subset S\vee S$, then the projection $\G_{W_0}$ onto
$W_0$ with the kernel $K$ defines an $\so(V)$-equivariant bracket
\begin{equation}\arr
[\ ,\ ] : &S\ot S \ra W_0 \\
          &[s,t] = \G_{W_0} (s\ot t)
\ea\label{bracket}\end{equation} which is skewsymmetric or symmetric,
respectively. More generally, if $A$ is an endomorphism of $W_0$ that
commutes with $\so(V)$, then the twisted projection $A \circ \G_{W_0}$ is
another $\so(V)$-equivariant bracket and any bracket can be obtained in
this way. Together with the action of  $\so(V)$ on  $W_0$ and $S$, this
defines the structure of an $\e$-transalgebra $\gg= \so(V) + W_0 + S$,
since the Jacobi identity for $X, Y , Z \in \gg_1$ follows from $[\gg_1 ,
[\gg_1 ,\gg_1] ] = 0$. The classification problem then reduces
essentially to the decomposition of $S\wedge S$, $S\vee S$ into
irreducible $\so(V)$-submodules and the description of the projection
$\G_{W_0}$. In this paper, we resolve both these matters. In all cases
the irreducible $\so(V)$-submodules occurring in the tensor product $S\ot
S$ are $k$-forms $\La^k V$, with the exception of the case of even
dimensions $n=p{+}q=2m$  with signature $s=p{-}q$ divisible by 4. In the
latter case the $m$-form module splits into irreducible selfdual and
anti-selfdual submodules $\La_\pm^m V$. The multiplicities of any
irreducible $\so(V)$-submodules of $S \ot S$ take values 1,2,4 or 8. For
example if $V=\bC^n\;,\; n=2m+1$ or if $V=\bR^{m,m+1}$, we have  (c.f.
\cite{OV})
\bean   S\ot S &=& \sum_{k=0}^{m} \La^k V\,, \\[10pt]
S\vee S &=& \sum_{k=0}^{[m/4]} \La^{m-4k}V
+ \sum_{k=0}^{[(m-3)/4]} \La^{m-3-4k}V
\,,
\\[10pt]
S\wedge S &=& \sum_{k=0}^{[(m-2)/4]} \La^{m-2-4k}V
                         +\sum_{k=0}^{[(m-1)/4]} \La^{m-1-4k}V \, .
\eean
The vector space of $\e{=}{-}1$-extensions of translational type of the form
$\gg= \so(V) + \La^kV + S$ is identified with the vector space
$\Bil^k_-(S)^{\so(V)} :=$ Hom$_{\so(V)}(S\wedge S\ ,\ \La^kV)$
of $\La^kV$-valued invariant skewsymmetric bilinear forms on $S$.
Similarly, the vector space of $\e{=}{+}1$-extensions of translational type
of the form $\gg= \so(V) + \La^kV + S$ is identified with the vector space
$\Bil^k_+(S)^{\so(V)} := $Hom$_{\so(V)}(S\vee S\ ,\ \La^kV)$.

The main problem is the
description of these spaces of invariant $\La^k V$-valued bilinear forms.
For $k = 0, 1$ this problem was solved in \cite{AC}, where three invariants,
$\s , \t$ and $\i$,  were defined for  bilinear forms on the spinor module.

Following \cite{AC}, a nondegenerate $\so(V)$-invariant (scalar) bilinear
form  $\beta$ on the spinor  module $S$ is called  {\it admissible} if it
has the following properties:
\begin{itemize}
\item[1)] $\beta$ is either symmetric or skewsymmetric,
$\b(s,t)= \sigma(\b)\b(t,s)\,,\, s,t\in S\,,\, \sigma(\b)=\pm 1$.
We  define $\sigma(\b)$ to be the {\it symmetry} of $\beta$.
\item[2)] Clifford multiplication by $v\in V$,
$$\g(v):S\ra S\, , \quad s\mapsto \g(v)s = v\cdot s\, ,$$
is either $\beta$-symmetric or $\beta$-skewsymmetric, i.e.\
$$\b(vs,t)= \tau(\b)\b(s,vt)\;,\quad s,t\in S\, ,$$
with
$\tau(\b)=+1$ or $-1$, respectively. We define  $\tau(\beta)$ to be the
{\it type} of $\beta$.
\item[3)] If the spinor module is reducible, $S = S^{+} + S^{-}$, then
the semispinor modules $S^+$ and $S^-$ are either mutually orthogonal or
isotropic. We define the  {\it isotropy} of $\beta$ to be $\i(\beta ) = +1$
if  $\b(S_+,S_-)=0$ or   $\iota(\beta ) = -1$ if $\b(S_\pm,S_\pm)=0$.
\end{itemize}
In \cite{AC}, a basis $\b_i$
of the space ${\rm Bil}(S)^{\so(V)} := {\rm Bil}^0(S)^{\so(V)}$ of
scalar-valued invariant forms was constructed
explicitly, which consists of admissible forms. These are tabulated in the
appendix (Table \ref{tbl:beta}).
The dimension
$N(s) = {\dim\ } {\Bil}(S)^{\so(V)}$ depends only on the signature  $s=p-q$
of $V$ (see Table \ref{tabA0} in the Appendix).  We associate with a
bilinear form $\b$ on $S$ the $\La^kV$-valued bilinear form
$\G^k_\b : S\ot S \ra \La^k V$, defined
by the
following fundamental formula
$$ \langle \G_\b^k (s\ot t)\,,\, v_1\wedge\dotsm\wedge v_k \rangle\
= \sum_{\pi\in \gS_k}  {\rm sgn}(\pi )
\b\left( \g(v_{\pi(1)})\dotsm\g(v_{\pi(k)}) s\,,\,t\right)\quad
s,t\in S\ ,v_i\in V\;,  $$
which extends the formula given in \cite{AC} from $k = 1$ to arbitrary $k$.
For $k=0$ we have that $\G^0_\b = \b$.

We shall prove that the
map $\beta \mapsto \G_\beta^k$ is $\so(V)$-equivariant and
induces an isomorphism
\[ \G^k : {\Bil}(S)^{\so(V)} \stackrel{\sim}{\ra}{\Bil}^k(S)^{\so(V)}\]
onto the vector  space of $\La^kV$-valued invariant bilinear forms
on $S$. This was proven for $k=1$ in \cite{AC}.

The definitions of the invariants $\s,\tau,\i$ make sense for
$\La^kV$-valued bilinear forms as well.
If $\s(\G^k_\b)=-1$, the form $\G^k_\b$ is skewsymmetric and hence defines a
Lie algebra structure on $\gg= \so(V) + \La^kV + S$. If $\s (\G^k_\b )= +1$,
it defines a Lie superalgebra structure on $\gg= \so(V)+\La^kV+S$.
We shall prove, for admissible $\b$, that
\begin{equation}
  \s(\G^k_\b) = \s(\b)\tau(\b)^k (-1)^{k(k-1)/2}\, .
 \label{sigmak}
\end{equation}
In the cases when semi-spinors exist, we shall prove that
\begin{equation}
  \i(\G^k_\b) = \i(\b)(-1)^k\, .
 \label{iotak}
\end{equation}
For $k>0$ the bilinear forms $\G_\beta^k$ associated with an
admissible bilinear form $\b$ have neither value of the
type $\tau$. Clearly, the formulae for the invariants show
that   $ \s(\G^k_\b)$ and $\i(\G^k_\b)$
depend only on $k$ modulo 4. We tabulate these  invariants for
$\G_{\beta_i}^k$ for $k=0,1,2,3$ in the Appendix.

Let the number $N^\e_k(s,n)$ denote the
dimension of the vector space  ${\Bil}_\e^k(S)^{\so(V)}$ of $\e$-extended
$k$-polyvector Poincar{\'e} algebra structures on
$\gg= \so(V) + \La^kV + S\;$. We shall see that the sum
$$N_k(s,n) = N^+_k(s,n)+ N^-_k(s,n) = N(s) = \dim\Bil(S)^{\so(V)}$$
depends only on the signature $s$.
We shall also verify the following remarkable shift
formula
\begin{equation}  N_k^\pm(s,n+2k) = N_0^\pm(s,n)\ ,
\label{shift}\end{equation} which reduces the calculation of these
numbers to the case of zero forms. The function $N^\pm(s,n):=
N_0^\pm(s,n)$ has the following symmetries:

\noindent
a) Periodicity modulo 8 in $s$ and $n$:
$$
 N^\pm(s+8a, n+8b) = N^\pm(s,n) \, ,\quad a,b\in\bZ\, .
$$
Using this, we can extend the functions $N^\pm(s,n)$ to all
integer values of $s$ and $n$. \\
b) Symmetry with respect to reflection in signature 3:
$$ N^\pm(-s+6,n) = N^\pm(s,n)\, .
$$
c) The mirror symmetries:
\bea
 N^\pm(s,n+4) &=& N^\mp(s,n)\,,
\label{mirror1}\\
 N^\pm(s,-n+4) &=& N^\mp(s,n)\, .
\label{mirror2}\eea
Due to the shift formula \re{shift}, all these identities yield
corresponding identities for $N_k^\pm(s,n)$ for any $k$.
For example the mirror identity \re{mirror2} gives the mirror symmetry
for $k{=}1$ (reflection with respect to zero dimension),
$$
N_1^\pm(s,-n) = N_1^\mp(s,n)\ ,
$$
which was discovered in \cite{AC}.

In Appendix~\ref{app:reformPhysicists}, we summarise our results in
language more familiar to the physics community.

Recently, there have been many discussions (e.g.\
\cite{AI,CAIP,DFLV,DN,FV,Sc,Sh,V,VV}) of generalizations of spacetime
supersymmetry algebras which go beyond Nahm's classification.  Of
particular interest, has been the M-theory algebra, which extends the
$d{=}11$ super Poincar{\'e} algebra by two-form and five-form brane charges.
In the important paper   \cite{DFLV}, the authors study superconformal
Lie algebras and polyvector super-Poincar{\'e} algebras $\gg = \so(V) +
\La^kV + W_1$, where $W_1 = S$ or $W_1 = S_\pm$. They propose an approach
for the classification of such Lie superalgebras  $\gg$ which consists
essentially of the following two steps: first describe the space
Hom$_{\so(V^{\bC})}(\bS\vee \bS \ ,\ \La^kV^{\bC})$, if the complex
spinor module $\bS$ is irreducible, and the spaces
Hom$_{\so(V^{\bC})}(\bS_\pm\vee \bS_\pm \ , \La^kV^{\bC})$ and
Hom$_{\so(V^{\bC})}(\bS_+\otimes \bS_- \ , \La^kV^{\bC})$ if the complex
spinor module $\bS = \bS_+ + \bS_-$ is reducible, then describe
$\so(V)$-invariant reality conditions. They determine the dimension of
the above vector spaces, which is always zero or one and discuss the
second problem.   In the present paper we start from the real spinor
module $S$ and, in particular, describe explicitly the real vector space
$H =$ Hom$_{\so(V)}(W_1\vee W_1 \ ,\La^kV)$ for an arbitrary spinorial
module $W_1$. We shall see that even if $W_1$ is an irreducible spinor
module $S$, the dimension of $H$ can be $0, 1, 2$ or  $3$. Polyvector
super-Poincar{\'e} algebras were also considered in \cite{CAIP} for
Lorentzian signature $(1,q)$ in the dual language of left-invariant
one-forms on the supergroup of supertranslations.

\section{$\boldsymbol{\e}$-extensions of $\boldsymbol{\so(V)}$}

Let $V$ be a real or complex vector space endowed with a scalar product
and $W_1$ an $\so(V)$-module.
\bd \label{superextDef}
A {\bss superextension ($\boldsymbol{\e = +1}$-extension) of
$\boldsymbol{\so(V)}$ of type $\boldsymbol{W_1}$}  is a Lie superalgebra
$\gg$   satisfying the conditions\\
i) $\so(V) \subset \gg_0$ as a subalgebra  \\
ii) $\,\gg_1 = W_1$, a $\,\gg_0$-module.\\
A {\bss Lie extension ($\boldsymbol{\e = -1}$-extension) of
$\boldsymbol{\so(V)}$ of type $\boldsymbol{W_1}$} is a $\bZ_2$-graded Lie
algebra $\gg =\gg_0 + \gg_1$, also satisfying i) and ii).
Further, an $\e$-extension is called {\bss minimal} if it does not
contain a proper subalgebra which is also an $\e$-extension of type $\,W_1$;
more precisely, if $\gg' = \gg_0'+\gg_1 \subset \gg$,
$\,\so(V)\subset \gg_0'$, then $\gg' = \gg$.
\ed

\noindent
The Lie superalgebras classified by Nahm are examples of superextensions
of $\so(\bR^{p,q})$ of spinor type $W_1=S$.

Let $\,\gg=\gg_0 + \gg_1$, be an $\e$-extension of $\so(V)$, $\,\gg_0=
\so(V) + W_0$, with $W_0$ an $\so(V)$-submodule that is complementary to
$\so(V)$ in $\gg_0$ and $\,\gg_1= W_1 $. There are two extremal classes
of such algebras:
\begin{itemize}
\item[E1:\,]  $\gg$  is semi-simple, i.e.\ does not
contain any proper solvable ideal,

\item[E2:\,]  $\gg$  is of {\it semi-direct type},
i.e.\ $\gg$  is maximally non semi-simple, in the sense that
$\so(V)$ is its largest semi-simple super Lie subalgebra,
$\,\gg= \so(V) + W_0 + W_1$
and $\,W= W_0 + W_1$ is a solvable ideal.

\end{itemize}

\section{Extensions of translational type and
$\boldsymbol{\e}$-transalgebras} \bd Let $\,\gg = \so(V) + W_0 + W_1 $ be
an  $\e$-extension of  $\so(V)$. If $[W_0 , W]=0$ and $[W_1,W_1] \subset
W_0$ then the extension $\gg = \so(V) + W_0 + W_1$ is called an {\bss
$\boldsymbol{\e}$-extension of $\boldsymbol{\so(V)}$ of translational
type} and the (nilpotent) ideal $\,W=  W_0 + W_1$ is called the {\bss
algebra of generalized translations}.  If it is minimal, in the sense of
Definition \ref{superextDef},  then it is called an {\bss
$\boldsymbol{\e}$-transalgebra}. \ed

We note that such an extension is automatically of semi-direct type,
provided that $\dim V \ge 3$, which we assume in this section. We also
assume for definiteness that $V$ is a {\it real} vector space. The
minimality  condition is equivalent to $[W_1,W_1] = W_0$ and means
that even translations are generated by odd translations.
The construction of $\e$-extensions of $\so(V)$ of translational type
with given $\so(V)$-modules $W_0$ and $W_1$ reduces to the
construction of $\so(V)$-equivariant linear maps $\vee^2 W_1 \ra W_0$ and
$\La^2W_1 \ra W_0$. The Jacobi identity for the Lie bracket associated
to such a map follows from the $\so(V)$-equivariance.
Now we show that the description of $\e$-extensions of $\so(V)$ of
translational type reduces to that of minimal ones (i.e.\ transalgebras).
Let $\gg = \so(V) + W_0 + W_1 $ be an  $\e$-extension of  $\so(V)$
of translational type. Then $\gg' := \so(V) + [W_1,W_1] + W_1 $ is
an $\e$-transalgebra and
$\gg = \gg' + \ga$ is the semi-direct sum of the subalgebra
$\gg'$ and an (even) Abelian ideal $\ga$, where $\ga \subset W_0$ is
an $\so(V)$-submodule complementary to $[W_1,W_1] \subset W_0$.
Conversely, if $\gg'= \so(V) + W_0' + W_1$
is an $\e$-transalgebra and $\ga$ is an $\so(V)$-module then
the semi-direct sum $\gg := \gg' + \ga$ is an $\e$-extension of  $\so(V)$
of translational type, where $W_0 := W_0'+ \ga$.

\bp Let $W_1$ be an $\so(V)$-module. Then there exists a unique (up to
isomorphism) $\e$-transalgebra of maximal dimension with $\gg_1 = W_1$:
\[ \gg^{\e} =  \gg^{\e}(W_1) = \gg^{\e}_0 + \gg^{\e}_1 =
(\so(V) + W_0^{\e}) + W_1\, ,\]
where  $W_0^+ = \vee^2W_1$ and  $W_0^- = \La^2W_1$. The Lie (super)bracket
$[\cdot ,\cdot ] : W_1\ot W_1 \ra W_0^{\e}$ is the projection
onto the corresponding summand of $W_1\ot W_1
= \vee^2W_1 \oplus \La^2W_1$. Moreover, any $\e$-transalgebra with
$\gg_1 = W_1$ is isomorphic to a contraction of $\gg^{\e}(W_1)$.
\ep

\pf It is clear that $\gg^{\e}$ is a maximal $\e$-transalgebra.
Let $\gg =  \gg_0 + \gg_1$  be a maximal $\e$-transalgebra with
$\gg_1 = W_1$ and $\gg_0 = \so(V) + W_0$.  The Lie (super)bracket
$[\cdot ,\cdot ] : W_1\ot W_1 \ra W_0$ defines an $\so(V)$-equivariant
isomorphism from $\vee^2W_1$ or $\La^2W_1$ onto $W_0$. This isomorphism
extends to an isomorphism  $\gg^{\e} \ra \gg$, which is the identity on
$\so(V) + W_1$. Similarly for any $\e$-transalgebra with $\gg_1 = W_1$
the (super) Lie bracket $[\cdot ,\cdot ] : W_1\ot W_1 \ra W_0$ defines an
$\so(V)$-equivariant epimorphism $\varphi$
from $\vee^2W_1$ or $\La^2W_1$ onto $W_0$.
The kernel $K$ is an $\so(V)$-submodule of $\vee^2W_1$ or $\La^2W_1$,
respectively. Since $\so(V)$ is semi-simple, there exists a
complementary submodule $\wt{W}_0$ isomorphic to $W_0$.
We can identify the $\so(V)$-module $\wt{W}_0$ with $W_0$ by means
of the isomorphism $\varphi|\wt{W}_0$. Then the Lie bracket corresponds
to the projection
$\pi^+_{\wt{W}_0}: \vee^2W_1 = K +  \wt{W}_0 \ra \wt{W}_0$ or
$\pi^-_{\wt{W}_0}:\La^2W_1 = K +  \wt{W}_0 \ra \wt{W}_0$.
This defines an
$\e$-transalgebra $\gg^{\e}(W_1,\wt{W}_0) = \so(V) + \wt{W}_0 +
W_1$, whose bracket is the above projection $\pi^{\e}_{\wt{W}_0}$.
The isomorphism $\varphi|\wt{W}_0 : \wt{W}_0 \ra W_0$ of
$\so(V)$-modules extends trivially to an isomorphism
$\gg^{\e}(W_1,\wt{W}_0) \ra \gg$. This shows that any $\e$-transalgebra
is isomorphic to an $\e$-transalgebra of the form $\gg^{\e}(W_1,\wt{W}_0)$,
where $\wt{W}_0 \subset W_1\ot W_1$   is an $\so(V)$-submodule contained
in $\vee^2W_1$ or  $\La^2W_1$ , respectively. Consider now the
one-parameter family of Lie brackets $[ \cdot ,\cdot ]_t :=
t({\rm Id} - \pi^{\e}_{\wt{W}_0}) +  \pi^{\e}_{\wt{W}_0}$. This
defines a family of $\e$-transalgebras $(\gg^\e(W_1),[ \cdot ,\cdot ]_t)$.
For $t\neq 0$ they are isomorphic to the original ($t=1$) $\e$-transalgebra.
In the limit $t \ra 0$ we obtain the $\e$-transalgebra
$\gg^{\e}(W_1,\wt{W}_0)$ as a contraction of $\gg^\e(W_1)$.
\qed

The following proposition describes the structure of extensions of
semi-direct type under the additional assumption that the $\so(V)$-module
$W_1$ is irreducible. We denote by $\rho: \gg_0 \ra \gg\gl(W_1)$ the
adjoint representation of $\gg_0= \so(V)+W_0$ on $W_1$ and by $K$ the
kernel of $\rho|_{W_0}$, which is clearly an ideal of $\gg$. Thus, $\rho
$ is the action of ${\rm ad}_{W_0}$ on $W_1$ and $K$ are all the
generators of $W_0$ that commute with $W_1$.

\bp
Let $\gg = \so(V) + W_0 + W_1 $ be an $\e$-extension of semi-direct type.
Assume that the  $\so(V)$-module $W_1$ is irreducible of dimension at least
3 if it does not admit  an $\so(V)$-invariant complex structure and
$\dim_\bC W_1 \ge 3$ if  it does and that $\dim V \ge 3$. Consider the
decomposition of $\gg$ into a direct  sum of $\so(V)$-submodules, $ \gg =
\so(V) + A + K +  W_1\,,$  where $A$ is an $\so(V)$-invariant complement to
$K$ in $W_0= A{+}K$.  Then the dimension  $\dim A=0,1,2$ and the irreducible
linear Lie algebra   $\rho(\gg_0) = \rho(\so(V)) + Z$, where the centre
$Z\cong W_0/K$ is either $0,\,\bR{\cdot}{\rm Id}\,$ or $\,\bC{\cdot}{\rm Id}$.
Moreover,
$$ [A,A] \subset K \quad,\quad [\so(V),A]=0\quad,\quad
[W_1,W_1]\subset K\ .  $$
\ep
\pf
By assumption, the  linear Lie algebra $\rho (\gg_0) \subset \gg\gl(W_1)$ is
irreducible and hence reductive. Since any solvable ideal of a reductive Lie
algebra belongs to the centre, we conclude that the solvable ideal
$\rho (W_0) \subset \rho (\gg_0)$ is in fact Abelian and consists of operators
commuting with $\so(V)$. Now Schur's Lemma implies that
$\rho(W_0) = 0,\,\bR{\cdot}{\rm Id}$, or
$\bC{\cdot}{\rm Id}$.  The inclusion  $ [A,A]
\subset K$  follows from the fact that $\rho(A)$ is in the centre of
$\gg_0$. Since the restriction of $\rho$ to $\so(V) + A$ is faithful and
$[\rho (\so(V)),\rho (A)] = 0$, we conclude that $[\so(V),A]=0$. From the
assumptions it follows that there exist three vectors $x,y,z \in W_1$,
which are linearly independent over the reals if  $W_1$ has no invariant
complex structure and over the complex numbers if  $W_1$ has an invariant
complex structure $J$. For any three linearly independent vectors (over
$\bR$ or $\bC$) $x,y,z \in W_1$, the Jacobi identity gives
$$\arr  0&=& [[x,y],z] + [[y,z],x] + [[z,x],y] \\
&=& \rho([x,y])z +   \rho([y,z])x +  \rho([z,x])y\ .
\ea$$
Since $[W_1,W_1]\subset W_0$ and $\rho(W_0)= \rho(A)$ consists of scalar
operators (over $\bR$ or $\bC$), we have that $\rho([x,y])=0$, i.e.\
$[W_1,W_1]\subset K= \ker\rho$.
\qed

\noindent
Note that $\gg$ is a transalgebra if and only if $A{=}0$.

The following corollary gives sufficient conditions for extensions
of semi-direct type to be transalgebras.

\bc
Under the assumptions of the previous proposition, assume moreover
that $\gg=\so(V){+}W_0{+}W_1 $ is a minimal extension of type $W_1$.
Then $\gg$ is a transalgebra.
\ec

\pf
Minimality implies $ W_0 = [W_1, W_1]$  and, by the above
Proposition,  $[W_1, W_1]$ commutes with $W_1$.
Now the Jacobi identity for $x,y \in W_1$ and $z \in W_0$ yields
$[W_0,W_0]=0$.

\qed

\noindent
Instead of minimality we may assume the irreducibility of the
$\so(V)$-module $W_0$.

\bp
Let $\gg = \so(V) + W_0 + W_1 $ be an $\e$-extension of semi-direct type,
with $\dim V \geq 3$.  Assume that $W_0$ and $W_1$ are irreducible
$\so(V)$-modules. Then either $\gg\,$ is of translational type, i.e.\
$[W_0,W]=0$, or  $W_0\cong \bR$ (considered as a real Lie algebra) is the
centre of  $\gg_0 = \so(V) + W_0\,$ and $\,\ad_{W_0}$ acts on $W_1$ by
scalars.
\ep
\pf
Let $W_0,W_1$ be irreducible $\so(V)$ modules.
Since the algebra $\,W=  W_0 + W_1$ is solvable,
$[W_0,W_0]$ is
a proper $\so(V)$ submodule of $W_0$, hence $[W_0,W_0] =0$.
The kernel $K$ of the adjoint representation $\rho: W_0 \ra \gg\gl(W_1)$
is an $\so(V)$-submodule of $W_0$. Hence $K=W_0$ or $0$. In the
first case, $\gg$  is of translational type. In the second case,
the representation $\rho$ is faithful and $\rho (W_0)$ commutes
with $\rho (\so(V))$, hence $[\so(V),W_0] = 0$. On the other hand the
$\so(V)$-module $W_0$ is irreducible, so $W_0\cong \bR$.
\qed
\goodbreak\newpage

\section[Extended polyvector Poincar{\'e} algebras]
{Extended polyvector Poincar{\'e} algebras
and $\boldsymbol{\La^kV}$-valued invariant bilinear forms on the spinor
module $\boldsymbol{S}$}
In this and the next two sections, we devote ourselves to the
classification of  $\e$-transalgebras $\gg= \gg_0 + \gg_1 $ with $\gg_1=
W_1=S$, the spinor  $\so(V)$-module. We take $V$ to be
the pseudo-Euclidean space $\bR^{p,q}$ of dimension $n=p{+}q$ and
signature $s=p{-}q$.  In other words, we consider (super) Lie algebras
$\gg = (\so(V) +W_0) + S$ with
$$
[W_0\ ,\ W_0+S ] =0\quad,\quad W_0 = [S\ ,\ S]\ .
$$
The (super) Lie bracket defines an $\so(V)$-equivariant surjective map
$\G_{W_0}: S\ot S\ra W_0$.
If $K$ is the kernel of this map, then  $S\ot S = \wt{W}_0+K$, where
$\wt{W}_0$ is an $\so(V)$-submodule equivalent to $W_0$ such that
$W_0 \subset S\wedge S$ in the Lie algebra case and
$W_0 \subset S\vee S$ in the superalgebra case.
Conversely, if we have a decomposition $S\ot S= W_0+K$ into a
sum of two $\so(V)$-submodules and moreover $W_0 \subset S\wedge S$
or $W_0 \subset S\vee S$, then the projection $\G_{W_0}$ onto
$W_0$ with the kernel $K$ defines an $\so(V)$-equivariant bracket
\begin{equation}\arr
[\ ,\ ] : &S\ot S \ra W_0 \\
          &[s,t] = \G_{W_0} (s\ot t)
\ea\end{equation} which is skewsymmetric or symmetric, respectively. More
generally, if $A$ is an endomorphism of $W_0$ which commutes with
$\so(V)$, then the twisted projection $A{\circ}\G_{W_0}$ is another
$\so(V)$-equivariant bracket and any bracket can be obtained in this way.
Together with the action of  $\so(V)$ on  $W_0$ and $S$, this defines the
structure of an $\e$-transalgebra $\gg= \so(V) + W_0 + S$. We therefore
have a 1--1 correspondence between $\e$-transalgebras of the form $\gg=
\so(V) + W_0 + S$, where $W_0$ is a submodule of $S\vee S$ (for $\e=1$)
or $S\wedge S$ (for $\e=-1$), and equivariant surjective maps $\G_{W_0}:
S\ot S\ra W_0$, whose kernel contains $S\vee S$ if $\e=-1$ and $S\wedge
S$ if $\e=1$. The problem thus reduces to the description of the
decomposition of $S\wedge S$ and $S\vee S$ into irreducible
$\so(V)$-submodules and the  determination of the twisted projections
$A{\circ}\G_{W_0}$. We consider these projections as equivariant
$W_0$-valued symmetric or skewsymmetric bilinear forms on $S$. In the
next section we show that the irreducible submodules of $S\ot S$ are of
the form $\La^k V$ or $\La_\pm^m V$ ((anti)selfdual $m$-forms) if $n=2m$
and $s$ is divisible by 4. We denote by
$$
{\rm Bil}^k (S) = {\rm Hom}(S\ot S\,,\,\La^kV)\ ,
$$
the vector space of $\La^k V$-valued bilinear forms on $S$.
It can be decomposed, ${\rm Bil}^k (S) = {\rm Bil}^k_+ (S)
\oplus {\rm Bil}^k_- (S)$, into the sum of the vector spaces of
symmetric ($+$) and skewsymmetric ($-$) bilinear forms.

For $W_0 = \La^k V$, the space of
$\e$-transalgebras ($\e = \pm$) is identified with the space
${\rm Bil}^k_\e (S)^{\so(V)}$
of ${\so(V)}$-invariant symmetric ($\e = +$) or skewsymmetric ($\e = -$)
bilinear forms.  Hence:\\
{\it The classification of $\e$-transalgebras $\gg = \gg_0 + \gg_1$ with
$\gg_1 = S$ reduces to the description
of the spaces ${\rm Bil}^k_\e (S)^{\so(V)}$ of $\La^k V$-valued invariant
bilinear forms on the spinor module $S$.}

The following formula associates a $\La^kV$-valued bilinear form
$\G_\b^k \in \Bil^k(S)$ to every (scalar) bilinear
form $\b\in {\rm Bil}(S)$.
$$
\langle \G_\b^k (s\ot t)\,,\, v_1\wedge\dotsm\wedge v_k \rangle\
=  \sum_{\pi\in \gS_k}
{\rm sgn}(\pi )\b\left(\g(v_{\pi(1)})\dotsm\g(v_{\pi(k)}) s\,,\,t\right)\quad
s,t\in S\ ,v_i\in V\;,
$$
where the sum is over permutations $\pi$ of $\{ 1,\dots,k\}$.
Our classification is based on the following theorem.
\bt \label{mainThm}
For any
pseudo-Euclidean vector space $V \cong \bR^{p,q}$, the map
$$\arr
\G^k : {\rm Bil}(S)  &\ra& {\rm Bil}^k (S)\\[8pt]
       \b &\mapsto& \G^k_\b
\ea$$
 is  a ${\rm Spin(V)}$-equivariant monomorphism
and it induces an isomorphism
$$
\G^k : {\rm Bil}(S)^{\so(V)} \stackrel{\sim}{\ra} {\rm Bil}^k (S)^{\so(V)}
$$
of vector spaces.
\et

\pf
It is known that Clifford multiplication $\g:V\ra {\rm End}\, S$ is
$\Spin(V)$--equivariant, i.e.\
$$
\g(gv) = g\g(v)g^{-1},\quad  g\in\Spin(V),\quad v\in V\ .
$$
Using this property we now check that the map $\G^k$ is also
$\Spin(V)$--equivariant: $$
\G_{g\cdot \beta}^k = g \cdot \G_\b^k\, ,
$$
where $(g\cdot \beta ) (s,t) = \beta (g^{-1}s,g^{-1}t)$ and
$(g \cdot \G_\b^k) (s,t) = g\G_\b^k(g^{-1}s,g^{-1}t)$.
We calculate
\bea
\langle  \G_{g\cdot \b}^k(s,t)\,,\, v_1\wedge\dotsm \wedge v_k \rangle
&=& \sum_{\pi\in \gS_k}
{\rm sgn}(\pi ) \b\left( g^{-1} \g(v_{\pi(1)})
\dotsm\g(v_{\pi(k)}) s\,,\,g^{-1}t\right)
\nonumber\\
&=& \sum_{\pi\in \gS_k} {\rm sgn}(\pi )
\b\left(\g(g^{-1}v_{\pi(1)})
\dotsm \g(g^{-1}v_{\pi(k)})g^{-1}s\,,\,g^{-1}t\right)
\nonumber\\
&=&\langle  \G_\b^k(g^{-1}s,g^{-1}t)\,,\,
g^{-1}v_1\wedge\dotsm \wedge g^{-1}v_k \rangle
\nonumber\\
&=&\langle  g\G_\b^k(g^{-1}s,g^{-1}t)\,,\, v_1\wedge\dotsm \wedge v_k \rangle
\nonumber\\
&=&\langle  (g\cdot \G_\b^k)(s,t)\,,\, v_1\wedge\dotsm \wedge v_k \rangle\ .
\eea
Next,  we prove that $\G$ is injective.
For $\b \in {\rm Bil}(S)$
the bilinear form $\G_\b^k$ is zero
if and only if
\[ \beta (\sum_\pi {\rm sgn}(\pi ) \g(v_{\pi(1)})\dotsm\g(v_{\pi(k)})S,S) = 0\]
or
\[ \sum_\pi {\rm sgn}(\pi ) \g(v_{\pi(1)})\dotsm\g(v_{\pi(k)})S \subset {\rm ker}
(\beta)\]
for any vectors $v_1,\dots,v_k$.
If the vectors $v_1,\dots,v_k$ are orthogonal, then the endomorphisms
$\g(v_1),\dots, \g(v_k)$ anticommute and the endomorphism
$\sum_\pi {\rm sgn}(\pi ) \g(v_{\pi(1)})\dotsm\g(v_{\pi(k)})=$  \linebreak
$k!\,\g(v_1)\dotsm\g(v_k)$ is invertible. This implies that
${\rm ker}(\beta) = S$ and so $\beta = 0$.

To complete the proof of the theorem, we need to check that
$$
\dim \Bil^k(S)^{\so(V)} = \dim \Bil(S)^{\so(V)} =: N(p-q)\, .$$
\noindent
In fact, $\dim \Bil^k(S)^{\so(V)} = \mu (k) \dim \cc (\La^kV)$, where
$\mu (k)$ is the multiplicity of $\La^kV$ in $S \otimes S$ and
$\cc (M) = {\rm End}_{\so(V)}(M)$ denotes the {\it Schur algebra} of
an ${\so(V)}$-module $M$. If the signature $s = p-q$ is divisible by 4
and $k = m = n/2$, then $\La^mV = \La_+^mV \oplus \La_-^mV$ is the
sum of two inequivalent irreducible ${\so(V)}$-modules of real type
and hence $\cc (\La^mV) \cong \bR \oplus \bR$. If the signature
$s$ is even but not divisible by 4
and $k = m = n/2$, then $\La^mV$ is an irreducible ${\so(V)}$-module
of complex type, with the complex structure defined by the Hodge star operator
and hence $\cc (\La^mV) \cong \bC$. In both cases
\[ \dim \Bil^m(S)^{\so(V)} = \mu (m) \dim \cc (\La^mV) = 2\mu (m) =
N(s)\, ,\]
where the last equation follows from Table \ref{tabA0} in the appendix.
In all other cases, $\La^kV$ is an irreducible module of real type and
$\cc (\La^kV) = \bR$. Therefore, using Table \ref{tabA0}, we obtain
\[ \dim \Bil^k(S)^{\so(V)} = \mu (k) \dim \cc (\La^kV) = \mu (k)
=   N(s)\, .\]
\qed

In the Introduction we defined the three $\bZ/2\bZ$-valued invariants for
$\La^k V$-valued bilinear forms on the spinor module: symmetry, type and
isotropy. We say that a non-zero $\La^kV$-valued bilinear form
$\G \in \Bil^k(S)$, $k>0$, is {\it admissible} if it is either
symmetric or skewsymmetric and, in the cases when semispinor modules
exist, if the two semispinor modules are either isotropic or mutually
orthogonal with respect to $\G$.
Recall that in the case of scalar-valued bilinear forms ($k=0$),
admissibility requires, in addition, that the bilinear form has a
specific type $\tau$.
The invariants of admissible  $\La^kV$-valued bilinear
forms in terms of the invariants of the scalar-valued admissible bilinear
forms are given by:
\bp \label{mainProp} Let $\beta \in \Bil(S)$ be a an admissible
scalar bilinear form and  $\G_\beta^k$ the associated  $\La^kV$-valued
bilinear form.  Then $\G_\beta^k$ is admissible and its invariants, the
symmetry $\s(\G^k_\b)$ and the isotropy $\i(\G^k_\b)$, can be calculated as
follows
\begin{eqnarray}
\s(\G^k_\b) &=& \s(\b)\tau(\b)^k (-1)^{k(k-1)/2}\, ,
\label{sigma}\\
\i(\G^k_\b) &=& \i(\b)(-1)^k\,.
\label{iota}
\end{eqnarray}
\ep
For $k>0$ the bilinear forms $\G_\beta^k \neq 0$ have neither type.

\pf
Let $s, t \in S$ and $e_1, \dots, e_k \in V$ be orthogonal vectors.
We put $\g_i := \g_{e_i}$ and compute
\begin{eqnarray*}
\langle \Gamma_\beta^k (s\otimes t), e_1\wedge \dotsm \wedge e_k\rangle &=&
k! \beta (\gamma_1 \cdots \gamma_k s, t)\\
&=& k! \t (\beta )^k \beta (s, \g_k \cdots \gamma_1t)\\
&=& k! \t (\beta )^k(-1)^{k(k-1)/2}\beta (s, \g_1 \cdots \gamma_kt)\\
&=& k! \t (\beta )^k(-1)^{k(k-1)/2}\s (\beta)\beta (\g_1 \cdots \gamma_kt,s)\\
&=& k! \t (\beta )^k(-1)^{k(k-1)/2}\s (\beta)
   \langle \Gamma_\b^k (t\otimes s), e_1\wedge \dotsm \wedge e_k\rangle \,.
\end{eqnarray*}
This proves equation (\ref{sigma}). Equation \re{iota} follows from the
fact that Clifford multiplication  $\g_v$ maps $S_+$ to $S_-$ and vice
versa.
\qed

\section[Decomposition of $\boldsymbol{S\ot S}$: complex case]
{Decomposition of the tensor square of the spinor module of
$\boldsymbol{\Spin(V)}$ into irreducible components: complex case}
\label{complex}

In this section we consider the spinor module $S$ of a complex Euclidean
vector space  $V= \bC^n$ and we derive the decompositions of $S\ot S$,
$S\vee S$ and  $S\wedge S$ into inequivalent irreducible
$\Spin(V)$--submodules. These decompositions also yield the corresponding
decompositions for the cases when $S$ is a  spinor module of a real
vector space $V= \bR^{m,m}$ if $n=2m$ and  $V= \bR^{m,m+1}$ if $n=2m+1$.
We shall use the well known facts summarised in the
following lemma, see e.g. \cite{OV}.

\bl
Let $V$ be an $n$-dimensional complex Euclidean vector space or a real
pseudo-Euclidean vector space of signature $(p,q)\ ,\ p+q{=}n\ ,\ p-q{=}s$.\\
If $n = 2m{+}1$, then the decomposition of $\La V$ into irreducible
pairwise inequivalent $\so(V)$-submodules is given by
\begin{equation}
\La V = \sum_{k=0}^n \La^k V = \sum_{k=0}^m \La^k V + \sum_{k=0}^m *\La^k V
= 2 \sum_{k=0}^m \La^k V \ .
\end{equation}
If $n=2m$ then we have the following decomposition into irreducible
pairwise inequivalent $\so(V)$-submodules
\bea
\La V = \begin{cases} 2 \displaystyle{\sum_{k=0}^{m-1}} \La^k V +\La^m V
&\text{if $s/2$  is odd}\\[15pt]
2\displaystyle{\sum_{k=0}^{m-1}} \La^k V +\La_+^m V + \La_-^m V
&\text{if $s/2$  is even or if $V$ is complex.}
\end{cases}\eea
Here $\La_\pm^m V$ are selfdual and anti-selfdual $m$-forms,
the $\pm 1$-eigenspaces of the Hodge $\ast$-operator,
 which acts isometrically on $\La^m V$,
with $\ast^2 = (-1)^{m+q} = (-1)^{s/2}= +1$ if $s/2$ is even .
\el
In particular, the  $\so(V)$-module $\La^k V$ is irreducible,
unless $n=2m$, $s/2$ is even and $k=m$, in which case
$\La^m V=\La_+^m V + \La_-^m V$,
where $\La_+^m
V$ and $\La_-^m V$ are irreducible inequivalent modules.

\bt\label{th_odd}
\begin{description}
\item{(i)} The $\Spin(V)$-module $S\ot S$ contains all modules $\La^kV$
which are irreducible.
\item{(ii)} If $V$ is a complex vector space of dimension $n=2m+1$ or
if $V$ is real of signature $(m,m+1)$  then
\bean
 S\ot S &=& \sum_{k=0}^{m} \La^k V
\,,
\\[10pt]
S\vee S &=& \sum_{i=0}^{[m/4]} \La^{m-4i}V
+ \sum_{i=0}^{[(m-3)/4]} \La^{m-3-4i}V
\,,
\\[10pt]
S\wedge S &=& \sum_{i=0}^{[(m-2)/4]} \La^{m-2-4i}V
                         +\sum_{i=0}^{[(m-1)/4]} \La^{m-1-4i}V \, .
\eean
\end{description}
\et

\pf (i) Theorem \ref{mainThm} associates a $\Spin(V)$-equivariant
linear map
\[ (\G_\b^k)^* : \La^kV \cong \La^kV^* \ra S^*\ot S^* \cong S\ot S\]
with any invariant bilinear form $\b$ on $S$. In particular, if
$ \La^kV$ is irreducible and $\b \neq 0$ then $(\G_\b^k)^*$ embeds
$\La^kV$ into $S\ot S$ as a submodule.   It was proven in \cite{AC} that
a non-zero invariant bilinear form $\b$ on $S$ always exists. This shows that
$S\ot S \supset  \sum_{k=0}^{m} \La^k V$.\\
(ii) If $n=2m+1$
then the right hand side has dimension $\frac{1}{2}2^n = 4^m$ and under
the assumptions on $V$ we have that $\dim S = 2^m$. Hence $\dim S\ot S = 4^m$,
so the inclusion is an equality. The decompositions of $S\vee S$ and
$S\wedge S$  can either be read off the tables in \cite{OV} or they follow
from Proposition \ref{mainProp} using the invariants
of the admissible scalar-valued  form, which in this case is unique up to
scale \cite{AC} (see the tables in  the  appendix).
\qed

Now, we consider the case when $V$ is complex of dimension $n=2m$ or real
of signature $(m,m)$. In this case, $\La^m V = \La_+^mV \oplus \La_-^mV$ and
the spinor module splits as a sum $S = S_+ + S_-$ of
inequivalent irreducible
semi-spinor modules $S_\pm$ of dimension $2^{m-1}$.

\bt\label{th_even}
 Let $V$ be complex of dimension $n=2m$ or real
of signature $(m,m)$. Then the
decompositions of  the $\Spin(V)$-modules
$S_+\ot S_-$ and $\ S_\pm \ot S_\pm\ $  into inequivalent
irreducible submodules are given by:
\bea
S_+ \ot S_- &=& \sum_{i=0}^{[(m-1)/2]} \La^{m-1-2i}V\,, \label{+-Equ}\\
S_\pm \ot S_\pm &=& \La_\pm^m V + \sum_{i=0}^{[(m-2)/2]} \La^{m-2-2i}V
\,,
\label{pmpmEqu}\\[8pt]
S\;\ot\; S\  &=& S_+ \ot S_+ + 2 S_+ \ot S_- + S_- \ot S_- = \La V\ .
\label{SSEqu}
\eea
Further, for any admissible bilinear form $\b$ on $S$,
the equivariant maps ${\G_\b^k|}_{S_\pm\ot S_\pm}$ and
${\G_\b^k|}_{S_+\ot S_-}$ have the following images:
\bea
&&\G_\b^m (S_\pm\ot S_\pm )\ =\ \La_\pm^m V \,,\label{proj1}\\[4pt]
&&\G_\b^{m\pm (2i+2)} (S_\pm\ot S_\pm )\
=\ \La^{m\pm (2i+2)}V\ ,\quad 0\le i\le [\fr{m-2}{2}]\,,
 \label{proj2}\\[4pt]
&&\G_\b^{m\pm (2i+1)}(S_+\ot S_- )\ =\ \La^{m\pm (2i+1)}V\ ,\quad
0\le i\le [\fr{m-1}{2}]\,,
 \label{proj3}\\[4pt]
&&\G_\b^{m\pm (2i+1)}(S_\pm\ot S_\pm )\ =\ 0\ ,\quad 0\le i\le [\fr{m-1}{2}]\,,
 \label{proj4}\\[4pt]
&&\G_\b^{m\pm 2i}(S_+\ot S_- )\ =\ 0\ ,\quad 0\le i\le [\fr{m}{2}]\ .
 \label{proj5}\eea
\et

\pf
To prove the theorem we use the following model for the spinor
module of an even dimensional complex Euclidean space $V$ or of a
pseudo-Euclidean space $V$ with split signature $(m,m)$: $V = U
\oplus U^*$, where $U$ is an $m$-dimensional vector space and the
scalar product is defined by the natural pairing between $U$ and the
dual space $U^*$. Then the spinor module is given by
$S = \La U= \La^{\rm ev}U + \La^{\rm odd}U = S_+ + S_-$,
where the semi-spinor modules $S_\pm$ consist of even and odd forms.
The Clifford multiplication is given by exterior and interior multiplication:
\begin{eqnarray*}
 u\cdot s &:=& u\wedge s \quad \mbox{for} \quad u\in U,\ s\in S\, ,\\
 u^*\cdot s &:=& \iota_u^*s\qquad \mbox{for} \quad u^*\in U^*,\ s\in S\, .
\end{eqnarray*}
There exist exactly two independent admissible bilinear forms $f$ and
$f_E=f(E\cdot\,,\,\cdot)$ on the spinor module,
where $E|_{S_\pm}=\pm {\rm Id}$, and the form $f$ is given by
\bea
f(\La^iU\,,\,\La^jU)  &=& 0\ ,\quad {\rm if}\quad i+j\neq m\,,
\nonumber\\
f(s,t) \vol_U &=& (-1)^{i(i+1)/2}\ s\wedge t\ ,\quad
s\in\La^iU\ ,\quad t\in\La^{m-i}U\ ,
\eea
where $\vol_U\in\La^mU$ is a fixed volume form of $U^*$.
We note that the symmetry, type and isotropy of the admissible
basis $(f,f_E)$ of $\Bil (S)^{\so(V)}$ are given by
\bean \s (f) = (-1)^{m(m+1)/2}\, ,\quad \s (f_E)
= (-1)^{m(m-1)/2}\, ,\\
 \t (f) = -1\, ,\; \t (f_E) =
+1\, ,\quad
 \i(f)= \i(f_E)= (-1)^m\, .
\eean
{}From this and Proposition \ref{mainProp} it follows that
\bea \label{sigmaffEEqu}
&\s (\G_f^k) = (-1)^{(m(m+1) + k(k+1))/2}\, ,\quad
\s (\G_{f_E}^k) = (-1)^{(m(m-1) + k(k-1))/2}\,,\\
&\i (\G_f^k) = \i (\G_{f_E}^k) = (-1)^{m+k}\, .
\eea
The formulae \re{proj2}-\re{proj5} and  the fact that
$\G_\b^m (S_\pm\ot S_\pm ) \neq 0$
follow from the formulae for the isotropy of $\G_f^k$ and $\G_{f_E}^k$.

To prove (\ref{proj1}) we  first show that for
any  admissible form
$\b$, the image
$\ \G_\b^m(S_+\ot S_+)\ $ contains $\ \La^m_+ V$ and the image
$\ \G_\b^m(S_-\ot S_-)\ $ does not contain $\ \La^m_+ V$.
For this we need to show that for any $a\in \La^m_+ V$ there exist spinors
$s,t\in S_+ = \La^{\rm ev}U$ such that the scalar product $\ \langle
\G_\b^m(s\ot t),a \rangle \neq 0\ $, and that there exists an element $a\in
\La^m_+ V$ such that $\langle \G_\b^m(s\ot t),a \rangle =0$ for any
$s,t\in S_-= \La^{\rm odd}U$. Since $\La_+V$ is an irreducible
$\so(V)$-module, it follows that if a single element $a$ of
$\La^m_+V$ is contained in the $\so(V)$-module $\Gamma_\beta  (S_+
\otimes  S_-)$, then all of $\La^m_+V$ is contained in it. Therefore, it
will suffice to prove the first statement for just one choice of $a$.

We shall use the following lemma. \bl Let $V = U \oplus U^*$ as above.
Then $\La^mU \subset \La^m_+V$. \el

\pf Let $(u_1,\dots ,u_m)$ be a basis of $U$ and $(u^*_1,\dots ,u^*_m)$ the
dual basis of $U^*$. Then, up to a sign factor, the volume form is given by
vol $=u_1\wedge \dots \wedge u_m\wedge u^*_1\wedge \dots \wedge u^*_m$.
Now, using the definition of the Hodge star operator,
$\langle \ast \a , \b \rangle {\rm vol} = \a \wedge \b$, we may immediately
check that  $\ast \left( u_1\wedge \dots \wedge u_m \right) = u_1\wedge \dots \wedge u_m$.
\qed

\noindent
Let us consider $a = {\rm vol}_U$. By the lemma,  $a \in \La^m_+V$.
Then for
$s=t=1\in S_+$ we have
$$ \langle \G_\b^m(s\ot t),a \rangle
= \b\left( a\wedge s\,,\,t \right) =  \b\left( a \,,\,1 \right)
= \pm 1 \neq 0\, .$$
Similarly, for any $s,t\in S_-$
$$ \langle \G_\b^m(s\ot t),a \rangle
= \b\left( a\wedge s\,,\,t \right) =  0 \, ,
$$
since deg$(a\wedge s)>m =\dim U$ and hence $a\wedge s=0$.
This proves both the above statements and hence $\G_\b^m (S_+\ot S_+ )
= \La_+^mV$. Since the image $\G^m_\b(S_-\ot S_-)$ is nonzero
and does not contain $\La^m_+V$, we also have
$\G^m_\b(S_-\ot S_-)=\La^m_-V$. This proves \re{proj1}.

We now prove \re{+-Equ}. By \re{proj3}, we have the inclusion
$\sum_{i=0}^{[(m-1)/2]} \La^{m-1-2i}V \subset S_+ \ot S_-$.
To prove equality we compare dimensions. Using the identity
$\;\binom{2m}{m-1-2i} = \binom{2m-1}{m-1-2i} + \binom{2m-1}{m-2-2i}\;$,
we calculate:
\bea
\dim \left(\sum_{i=0}^{[(m-1)/2]} \La^{m-1-2i}V \right)
&=& \sum_{i=0}^{[(m-1)/2]}\binom{2m}{m-1-2i}
= \sum_{i=0}^{m-1}\binom{2m-1}{i} \\
&=& \fr12 \sum_{i=0}^{2m-1}\binom{2m-1}{i}
= 2^{2m-2} =  \dim (S_+ \ot S_-)
\eea
since $\dim S_\pm =  2^{m-1}$.
This proves \re{+-Equ}.

Similarly, by \re{proj1} and \re{proj2}, we have
$S_\pm\ot S_\pm \supset \La^m_\pm V + \sum_{i=0}^{[(m-2)/2]}\La^{m-2i-2}V$.
To prove \re{pmpmEqu} we compare dimensions:
\bean  \dim  \left( \La_\pm^m V + \sum_{i=0}^{[(m-2)/2]} \La^{m-2i-2}V \right)
&=& \sum_{i=0}^{[m/2]}\binom{2m}{m-2i} -\frac{1}{2}\binom{2m}{m}\\
&=& \sum_{i=0}^{[m/2]}\left( \binom{2m-1}{m-2i}
   + \binom{2m-1}{m-2i-1}\right)  -\frac{1}{2}\binom{2m}{m}\\
&=& \sum_{i=0}^m\binom{2m-1}{m-i} -\frac{1}{2}\binom{2m}{m}\\
&=& \frac{1}{2}\sum_{i=0}^{2m-1} \binom{2m-1}{i}
= 2^{2m-2} = 2^{m-1}\cdot 2^{m-1} \\
&=& \dim (S_\pm \ot S_\pm )\, .
\eean
This proves \re{pmpmEqu} and \re{SSEqu}. \qed

\bc
\begin{enumerate}
\item[(i)]
Let $V$ be either complex of even dimension or real of signature $(m,m)$
and $\b$ an admissible
bilinear form on the spinor module $S = S_+ + S_-$.
Then for all $k$ the image of $\G_\b^k$ restricted
to $S_+ \ot S_+$, $S_-\ot S_-$ and $S_+\ot S_-$ is
an  irreducible $\Spin(V)$-module and the $\Spin(V)$-module $S\ot S$ is
isomorphic to $\La V$.
\item[(ii)] Let $V$ be either complex of odd dimension or real of
signature $(m,m+1)$ and $\b$ an admissible
bilinear form on the spinor module $S$.
Then for all $k$ the image $\G_\b^k (S\ot S)$ is irreducible
and the $\Spin(V)$-module $2S\ot S$ is isomorphic to $\La V$.
\end{enumerate}
\ec

\bc \label{cor_even}
 Let $V$ be complex of dimension $n=2m$ or real
of signature $(m,m)$. Then we have
\bea
S_\pm \vee S_\pm &=& \La_\pm^m V + \sum_{i=0}^{[(m-4)/4]} \La^{m-4-4i}V\,,
\\[10pt]
S_\pm \wedge S_\pm &=& \sum_{i=0}^{[(m-2)/4]} \La^{m-2-4i}V\ .
\eea
\ec

\pf These decompositions follow from \re{pmpmEqu} and \re{sigmaffEEqu}.
\qed

\section[Decomposition of $\boldsymbol{S\ot S}$:  real case]
{Decomposition of the tensor square of the spinor module of
$\boldsymbol{\Spin(V)}$ into  irreducible components: real case}

In this section we describe the decompositions of $S\ot S$, $S\vee S$ and
$S\wedge S$ into inequivalent irreducible $\Spin(V)$--submodules, where
$S$ is the spinor module of a pseudo-Euclidean vector space $V=\bR^{p,q}$
of arbitrary signature $s=p-q$ and dimension $n = p+q$. We obtain these
decompositions in two steps: First, we describe the complexification
$S^\bC$ of the spinor module $S$. Second, using the decomposition of the
tensor square $\bS \ot \bS$ of the complex spinor module $\bS$, we
decompose $S^\bC\ot S^\bC$ into complex irreducible
$\Spin(V^\bC)$--submodules and then we take real forms. We recall that
the complex spinor module $\bS$ associated to the complex Euclidean space
$\bV= V^\bC = V\ot \bC = \bC^n$ is the restriction to ${\rm Spin}(\bV)$
of an irreducible representation of the complex Clifford algebra $\Cl
(\bV)$.

Depending on the signature $s\equiv p-q\!\pmod 8$,
the complexification $S^\bC$ of the spinor module $S$ is given by either

\noindent
i) $S^\bC = \bS$, where we denote by $\bS$ the spinor module
of the complex Euclidean space $\bV= V^\bC = V\ot \bC = \bC^n$, or

\noindent
ii) $S^\bC = \bS + \bbS$, where $\bbS$ is the complex conjugated module
of $\bS$.

\noindent
In the latter case $S$ admits a $\Spin(V)$-invariant complex structure $J$
and $\bS$ is identified with the complex space $(S,J)$ and
$\bbS$ with $(S,-J)$. In the next lemma we specify the signatures for which
the cases i) or ii) occur. For this we use Table \ref{TabCl}, in which
we have collected important information about the real and complex Clifford
algebras  and spinor modules.

Now, we define the notion of Type$_{\Cl^0 (V)}(\bS,\bS_\pm)$
used in Table  \ref{TabCl}. If $s$ is odd, then the complex spinor module
$\bS$ is irreducible (as a complex module of the real even Clifford algebra
$\Cl^0 (V)$).   In this case we define
Type$_{\Cl^0 (V)}(\bS ) := \bK \in \{ \bR , \bH \}$ if the $\Cl^0 (V)$-module
$\bS$ is of real or quaternionic type, i.e.\ it admits a real or
quaternionic structure commuting with $\Cl^0 (V)$. For even $s$ the
complex spinor module $\bS = \bS_+ + \bS_-$ and $\bS_\pm$ are
irreducible complex $\Cl^0 (V)$-modules. We put
Type$_{\Cl^0 (V)}(\bS , \bS_\pm ) = (l\bK , \bK')$, where
$\bK$ and $\bK'$ are the types of $\bS$ and $\bS_\pm$, respectively, further
$l = 1$ if $\bS$ is irreducible and $l = 2$  if  $\bS_+$ and $\bS_-$
are not equivalent as complex  $\Cl^0 (V)$-modules. Note that if
the semispinor modules are of complex type ($s = 2, 6$), then
they are complex-conjugates of each other: $\ol{\bS}_\pm \cong
\bS_\mp$. If $\bS_\pm$ are of real ($s=0$) or quaternionic ($s=4$) type,
then they are selfconjugate:  $\ol{\bS}_\pm \cong \bS_\pm$.

We now explain how Table \ref{TabCl} has been obtained. The first two
columns have been extracted from \cite{LM} and imply the third column.
Passing to the complexification of the Clifford algebras we have:
$\Cl (V)\ot \bC = \Cl (V\ot \bC )$ and
$\Cl^0 (V)\ot \bC = \Cl^0 (V\ot \bC )$. {}From this we can describe
the complex spinor module $\bS$ and semispinors modules $\bS_\pm$ and
determine the relation between $S$, $S_\pm$  and $\bS$, $\bS_\pm$.
This gives the fourth, fifth and sixth columns of the table.
Using this table we prove the following lemma, which
describes the complex $Spin(\bV)$--module $S^\bC$:
\bl\label{lemSC}
$$
S^\bC= \left\{ \begin{array}{ll}
\bS + \ol{\bS} \quad&if\ s=p-q\equiv 1,2,3,4,5\!\pmod 8\\
\bS \quad&if\  s\equiv 6,7,8\!\pmod 8\ .
\ea\right.$$
\el
\pf
According to Table \ref{TabCl}, if $s\equiv 6,7,8\!\pmod 8$
we have $\bS =S\ot \bC$. In all other cases there exists a
$\Spin(V)$-invariant complex structure $J$ and  the complex space $(S,J)$ is
identified with $\bS$. Then $S^\bC= \bS + \bbS$.
\qed

\noindent
{\bf Remark 1:}
We note that a
$\Spin(V)$--invariant real or quaternionic structure $\f$  on $\bS$ (i.e.\ an
antilinear map with $\f^2= +1$ or $-1$, respectively) defines an isomorphism
$\f:\bS\ra\bbS$.  From Table \ref{TabCl} it follows that if $s\equiv
1,2\!\pmod 8$, then there exists a $\Spin(V)$--invariant real structure and
if $s\equiv 3,4,5\!\pmod 8$, then there exists a $\Spin(V)$--invariant
quaternionic structure on $\bS$.
\begin{table}\begin{center}
\begin{tabular}
{|c   |c        |c      |c      |c      |c          |c   |c   | } \hline
 $s$ & $\Cl_{p,q}$& $\Cl_{p,q}^0$& $\cc$ & $\bS$ & $\bS_\pm$ &
Type$_{\Cl^0 (V)}(\bS,\bS_\pm)$ & Name  \\ \hline
0 & $\bR(N)$ & $2\bR(N/2)$ & $2\bR $& $S\ot \bC $ & $S_\pm\ot \bC $ &
$(2\bR ,\bR)$& M-W \\
1& $\bC(N)$  & $\bR(N)$   &$\bR(2)$  & $S = S_\pm \ot \bC$ &  &$\bR$& M\\
2& $\bH(N/2)$& $\bC(N/2)$ & $\bC(2)$ & $S = S_\pm \ot \bC$ &
$S_\pm$ & $(2\bC ,\bC)$& M, W\\
3& $2\bH(N/2)$& $\bH(N/2)$& $\bH$ & $S$ &   &  $\bH$ & Symp\\
4& $\bH(N/2)$ &$2\bH(N/4)$&$2\bH$ & $S$ &$S_\pm$ & $(2\bH ,\bH)$& Symp-W \\
5& $\bC(N)$ & $\bH(N/2)$&$\bH$ & $S$ &   &  $\bH$ & Symp\\
6 & $\bR(N)$ & $\bC(N/2)$ & $\bC$ & $S\ot \bC $ & $S$ & $(\bR,\bC)$& M, W\\
7& $2\bR(N)$& $\bR(N)$&$\bR$& $S\ot \bC $ & &  $\bR$& M\\
\hline
\end{tabular}
\caption{\it Clifford Modules $\Cl_{p,q}$, their even parts $\Cl_{p,q}^0$,
the Schur algebra $\cc= \End_{\Cl^0(V)}(S)$, the complex Spinor Module
$\bS$, the complex Semispinor Modules $\bS_\pm$, the Type of these
$\Cl^0(V)$-modules and physics terminology: M stands for Majorana, W for
Weyl and  Symp for symplectic (i.e.\ quaternionic) spinors;
$s=p-q\mathop{\rm mod}\nolimits 8\;,\;n=p+q\;,\;N=2^{[n/2]}$. Note that
$p$ is the number of negative eigenvalues of the product of two gamma
matrices, and $q$ the number of positive eigenvalues, see
appendix~\ref{app:ClComReal}.} \label{TabCl}
\end{center}\end{table}

\noindent
Now, using the results of the previous section for the complex case,
we decompose $S^\bC\ot S^\bC$ into complex irreducible
$\Spin(\bV)$--submodules. If $S^\bC\ot S^\bC= \sum \bW_i$ and all
submodules $\bW_i$ are of real type  (i.e.\ complexifications of irreducible
real $\Spin(V)$--submodules $W_i$), then $S\ot S= \sum W_i$ is the desired
decomposition. In  odd dimensions all modules $\bW_i=\La^i \bV$ are
of real type. This is also the case in even dimensions $n=2m$,
with one
exception: the modules  $\La^m_\pm \bV\subset S^\bC\ot S^\bC$
are not of real type if
$*^2=-1$, i.e.\ if $s/2$ is odd. Then, $\La^m \bV = \La^m_+\bV + \La^m_-\bV$
is the complexification of the irreducible
$\Spin(V)$--module $\La^m V$, which has the $\Spin(V)$--invariant complex
structure $*$.

The decompositions of $S\vee S$ and $S\wedge S$ can be obtained using the
same method. Using this approach, we describe in detail all these
decompositions for any pseudo-Euclidean vector space $V = \bR^{p,q}$ in the
next three subsections.
\goodbreak
\subsection{Odd dimensional case: $\dim V=2m+1$}

We now describe the decomposition of $S\ot S = S\vee S + S\wedge S$ for all
signatures $s=1,3,5,7\!\pmod 8$ in the odd dimensional case.

\bt
Let $V=\bR^{p,q}$ be a pseudo-Euclidean vector space of dimension
$n=p+q=2m{+}1$. Then the decompositions of the $\Spin(V)$-modules $S\ot S$,
$S\vee S$ and  $S\wedge S$ into inequivalent irreducible submodules is given
by the following:

\noindent
If the signature $\ s=p{-}q\equiv 1,3,5\!\pmod 8\ $, we have
\bea
S\ot S &=& 2 (\La V) =
4 \sum_{i=0}^m \La^i V\,,
\label{decompa}\\[6pt]
\nonumber S \vee S &=& 3\sum_{i=0}^{[m/4]} \La^{m-4i}V
  + 3\sum_{i=0}^{[(m-3)/4]} \La^{m-3-4i}V \\
  && +\sum_{i=0}^{[(m-2)/4]} \La^{m-2-4i}V
  + \sum_{i=0}^{[(m-1)/4]} \La^{m-1-4i}V\,,
\\[6pt]
\nonumber S\wedge S &=& 3\sum_{i=0}^{[(m-2)/4]} \La^{m-2-4i}V
  + 3\sum_{i=0}^{[(m-1)/4]} \La^{m-1-4i}V \\
  && +\sum_{i=0}^{[m/4]} \La^{m-4i}V
  + \sum_{i=0}^{[(m-3)/4]} \La^{m-3-4i}V    \, .
\eea
If the signature $\ s\equiv 7\!\pmod 8\ $, we have
\bea
S\ot S &=& \sum_{i=0}^m \La^i V \,,
\label{decomp1}\\[6pt]
S\vee S &=& \sum_{i=0}^{[m/4]} \La^{m-4i}V
+ \sum_{i=0}^{[(m-3)/4]} \La^{m-3-4i}V\,,
\\[6pt]
S\wedge S &=& \sum_{i=0}^{[(m-2)/4]} \La^{m-2-4i}V
+ \sum_{i=0}^{[(m-1)/4]} \La^{m-1-4i}V\ .
\eea
Moreover, $S$ is an irreducible $\Spin(V)$-module for $s=3,5,7$
and for $s=1$, $S= S_+ + S_-$ is the sum of two equivalent semi-spinor
modules.
\et

\pf
The signature $s\equiv 7\!\pmod 8$ corresponds to $\bR^{m,m+1}$, which
was already discussed in Theorem \ref{th_odd}.

For  $s\equiv 1,3,5\!\pmod 8$, the spinor module $S$ has an invariant
complex structure $J$ and $(S,J)$ is identified with
the complex spinor module $\bS$.
We denote by $\bbS= (S,-J)$ the module conjugate to $\bS$. According to
Lemma \ref{lemSC} and Remark 1, $\bS$ and $\bbS$ are equivalent
as complex modules of the real
spin group $\Spin(V)$ and $\bS^\bC =  \bS + \bbS = 2\bS$. Hence,
\begin{eqnarray*}
 (S\ot_{\bR} S)^\bC &=&  S^\bC\ot_\bC S^\bC =  4 \bS\ot_\bC \bS\,, \\
(\vee^2  S)^\bC &=& \vee^2(\bS + \bbS)= \vee^2(2\bS ) =
 3\vee^2 \bS + \La^2 \bS\,, \\
(\La^2 S)^\bC &=& 3\La^2\bS + \vee^2 \bS \, .
\end{eqnarray*}
This implies the theorem. For example,
$$ (S\ot_\bR S)^\bC = 4 \bS\ot_\bC \bS = 4\sum_{i=0}^m \La^i \bV
= 4\sum_{i=0}^m (\La^i V)^\bC\, ,
$$
in virtue of Theorem \ref{th_odd} and the real part gives \re{decompa}.
\qed

\subsection{Even dimensional case: $\dim V=2m$}

In this subsection, we describe the decomposition of
$S\ot S = S\vee S + S\wedge S$ for all signatures $s=0,2,4,6\!\pmod 8$
in the even dimensional case.

\bt
Let $V=\bR^{p,q}$ be a pseudo-Euclidean vector space of dimension $n=p+q=2m$.
Then the decompositions of the $\Spin(V)$-module $S\ot S$ into inequivalent
irreducible submodules is given by  the following:

\noindent
If the signature $\ s=p{-}q\equiv  2,4\!\pmod 8$, we have
\begin{equation}
S\ot S =   4(\La V) = \begin{cases}
8\displaystyle{\sum_{i=0}^{m-1}} \La^i V + 4\La^m V
&\text{if}\  s=2\!\pmod 8  \\[14pt]
8\displaystyle{\sum_{i=0}^{m-1}} \La^i V + 4\La^m_+ V +4\La^m_- V
&\text{if}\  s=4\!\pmod 8.
\end{cases}\end{equation}
If the signature $\ s\equiv 0,6\!\pmod 8\ $, we have
\begin{equation}
S\ot S=  \La V = \begin{cases}
2\displaystyle{\sum_{i=0}^{m-1}} \La^i V + \La^m_+ V +\La^m_- V
&\text{if}\  s=0\!\pmod 8   \\[14pt]
2\displaystyle{\sum_{i=0}^{m-1}} \La^i V + \La^m V
&\text{if}\  s=6\!\pmod 8.
\end{cases}\label{decomp2}\end{equation}
\et
\pf
Similarly to the odd dimensional case, we have
\begin{equation} (S\ot_\bR S)^\bC = \bS\ot_\bC\bS  = \La \bV
= 2 \sum_{i=0}^m \La^i \bV + \La^m_+ \bV +\La^m_- \bV
\,.
\label{decomp5}\end{equation}
Now, we note that $\left(\La^m V\right)^\bC
=  \La^m \bV$ and $*^2|_{\La^mV}= (-1)^{s/2}$. If $s/2$ is even, then
$\La^mV= \La^m_+ V +\La^m_- V$, where $\La^m_\pm V$ are irreducible
submodules, which are $\pm 1$--eigenspaces of $*$. Then, $\left(\La^m_\pm
V\right)^\bC = \La^m_\pm \bV$. In this case, the real part of
\re{decomp5} gives the first part of \re{decomp2}. If $s/2$ is odd, then
$*$ is a complex structure on $\La^m V$, which is irreducible since it
has complex structure and its complexification $\La^m\bV$ has only two
irreducible components $\La^m_\pm\bV$. In this case the real part of
\re{decomp5} gives the second part of \re{decomp2}. \qed

The decompositions of $S\vee S$ and $S\wedge S$ for the cases when
semispinor modules exist, in particular for $s=0,2,4\!\pmod 8$,
will be given in the next subsection. Therefore it is now sufficient to
determine these decompositions for $s=6\!\pmod 8$.
\bc
\label{cor4}
 Let $S$ be the spinor module of a pseudo-Euclidean vector
space $V$ of signature $s\equiv 6\!\pmod 8$ and dimension $n = 2m$.
Then we have
\bea
S \vee S&=& \La^m V + 2\sum_{i=0}^{[(m-4)/4]} \La^{m-4-4i}V +
\sum_{i=0}^{[(m-1)/2]} \La^{m-1-2i}V\,, \\[10pt]
S \wedge S &=& 2\sum_{i=0}^{[(m-2)/4]} \La^{m-2-4i}V +
\sum_{i=0}^{[(m-1)/2]} \La^{m-1-2i}V \ .
\eea
\ec

\pf This follows by complexification of equation \re{decomp2},
using Lemma  \ref{lemSC}, equation \re{+-Equ} and Corollary \ref{cor_even}.
\qed

\subsection{Decomposition of tensor square of semi-spinors}

According to Table \ref{TabCl}, semi-spinor modules $S_\pm$ exist if the
signature $s\equiv 0,1,2,4\!\pmod 8$. More precisely, we list below
whether $S_\pm$ are equivalent Spin($V$)--modules and we give
$S_\pm^\bC\,$.

\begin{center}
\begin{tabular}
{|c     |c       |c          | }
\hline
 s  & $S_\pm$  & $S_\pm^\bC$\cr
\hline
0 & inequivalent & $S_\pm\ot \bC= \bS_\pm $\\
1 & equivalent & $S_\pm\ot \bC= \bS$\\
2 &equivalent & $S_\pm\ot \bC= \bS$\\
4 & inequivalent & $\bS_\pm + \bbS_\pm = 2 \bS_\pm$\\
\hline
\end{tabular}\end{center}

\noindent
For $s\equiv 1,2,4\!\pmod 8$ we have $\bS= S$, whereas for
$s\equiv 0\!\pmod 8$ we have $\bS= S^\bC$. We also note that for
$s\equiv 2\!\pmod 8$, although the Spin($V$)--modules $S_\pm$ are
equivalent, the Spin($\bV$)--modules $\bS_+$ and $\bS_-= \bbS_+$ are not
equivalent. For $s=4$, $S_\pm = \bS_\pm$ admits a $\Spin(V)$-invariant
quaternionic structure.

Using the above description for $S_\pm^\bC$ and the decompositions of
the tensor
squares of complex spinor and semi-spinor modules, we obtain the following:

\bt \label{thm:6} Let $S=S_+ + S_-$ be the spinor module of a
pseudo-Euclidean vector space $V$ of signature $s\equiv 0,1,2,4\!\pmod 8$
and dimension $n=2m$ or $n=2m+1$. Then we have the following
decomposition of $\Spin(V)$--modules $S_\pm\ot S_\pm$ and $S_+\ot S_-$
into inequivalent irreducible submodules:

\noindent
For $s\equiv 0\!\pmod 8$:
\bean
 S_+ \ot S_- &=& \sum_{i=0}^{[(m-1)/2]} \La^{m-1-2i}V\,,
\\[6pt]
 S_\pm \ot S_\pm &=& \La_\pm^m V + \sum_{i=0}^{[(m-2)/2]} \La^{m-2-2i}V\,,
\\[6pt]
S_\pm \vee S_\pm &=& \La_\pm^m V + \sum_{i=0}^{[(m-4)/4]} \La^{m-4-4i}V\,,
\\[6pt]
S_\pm \wedge S_\pm &=& \sum_{i=0}^{[(m-2)/4]} \La^{m-2-4i}V\, .
\eean
For $s\equiv 1\!\pmod 8$:
\bean
S_\pm \ot S_\pm &=& S_+\ot S_- =  \sum_{i=0}^{m} \La^i V\,,
\\[6pt]
S_\pm\vee S_\pm &=& \sum_{i=0}^{[m/4]} \La^{m-4i}V + \sum_{i=0}^{[(m-3)/4]}
\La^{m-3-4i}V \,,
\\[6pt]
S_\pm \wedge S_\pm &=& \sum_{i=0}^{[(m-2)/4]} \La^{m-2-4i}V
                         +\sum_{i=0}^{[(m-1)/4]} \La^{m-1-4i}V\, .
\eean
For $s\equiv 2\!\pmod 8$:
\bean
 S_\pm \ot S_\pm &=& S_+\ot S_-
= \La^m V + 2\sum_{i=0}^{m-1} \La^{i}V
 \,,
 \\[6pt]
S_\pm \vee S_\pm &=& \La^m V
+ 2 \sum_{i=0}^{[(m-4)/4]} \La^{m-4-4i}V +\sum_{i=0}^{[(m-1)/2]}
\La^{m-1-2i}V\,,
\\[6pt]
S_\pm \wedge S_\pm &=&
2\sum_{i=0}^{[(m-2)/4]} \La^{m-2-4i}V
                         +\sum_{i=0}^{[(m-1)/2]} \La^{m-1-2i}V\, .
\eean
For $s\equiv 4\!\pmod 8$:
\bean
 S_+ \ot S_- &=& 4\sum_{i=0}^{[(m-1)/2]} \La^{m-1-2i}V\,,
\\[6pt]
 S_\pm \ot S_\pm &=& 4\La_\pm^m V + 4\sum_{i=0}^{[(m-2)/2]} \La^{m-2-2i}V\,,
\\[6pt]
S_\pm \vee S_\pm &=& 3\La_\pm^m V + 3\sum_{i=0}^{[(m-4)/4]} \La^{m-4-4i}V
+\sum_{i=0}^{[(m-2)/4]} \La^{m-2-4i}V\,, \\[6pt]
S_\pm \wedge S_\pm &=& 3\sum_{i=0}^{[(m-2)/4]} \La^{m-2-4i}V
+\La_\pm^m V + \sum_{i=0}^{[(m-4)/4]} \La^{m-4-4i}V\ .
\eean
\et

\pf
The case $s=0\!\pmod 8$ follows from the complex case
(see Theorem \ref{th_even} and Corollary \ref{cor_even}).
For $s=1,2\!\pmod 8$ the modules $S_+$ and
$S_-$ are isomorphic. Hence $S = S_+ + S_- = 2 S_+$ and $S\ot S= 4 S_+\ot
S_+$. Since $S\ot S= 4 \sum_{i=0}^{m} \La^i V$ we have
$$
S_+\ot S_+ = S_-\ot S_- = S_+\ot S_- = \sum_{i=0}^{m} \La^i V\ .
$$
The splitting into symmetric and skew parts of these tensor products
follows that in the complex cases, see Theorem \ref{th_odd}
and Corollary \ref{cor_even}.
For  $s=4\!\pmod 8$ the semi-spinor modules $S_\pm$ are not equivalent but
$S_{\pm}^\bC = \bS = \bS_+ + \bS_-$. The result follows
from the decomposition in the complex case (Theorem \ref{th_even} and
Corollary \ref{cor_even}) on taking the real parts.
\qed

\section{$\boldsymbol{N}$-extended polyvector Poincar{\'e} algebras}
\label{N-extention}

In the previous sections we have classified $\e$-transalgebras of the
form $\gg = \gg_0 + \gg_1$, $\gg_0 = \so(V) + W_0$, with $\gg_1 = W_1 =
S$ the spinor module. As in \cite{AC} we can easily extend this
classification to the case where $W_1$ is a general spinorial module,
i.e.\ $W_1 = NS = S \otimes \bR^N$, if $S$ is irreducible, or $W_1 =
N_+S_+  \oplus N_-S_- = S_+ \otimes \bR^N_+ \oplus S_- \otimes \bR^N_-$
if semi-spinors exist. An $\e$-extensions of translational type of the
above form is called an {\bss $\boldsymbol{N}$-extended polyvector
Poincar{\'e} algebra} if $W_1 = NS$ and  an {\bss
$\boldsymbol{(N_+,N_-)}$-extended polyvector Poincar{\'e} algebra} if $W_1 =
N_+S_+  \oplus N_-S_-$. Consider first the case $W_1 = NS$.  As before,
the classification reduces to the decomposition of $\vee^2W_1$ and
$\La^2W_1$ into irreducible submodules. These decompositions follow from
the decompositions of $\vee^2S$ and  $\La^2S$ obtained in the previous
sections, together with the decompositions
\begin{eqnarray*}
\vee^2W_1 &=&
\vee^2S\otimes \vee^2 \bR^N \oplus   \La^2S\otimes \La^2\bR^N\,, \\
\La^2W_1 &=& \La^2S\otimes \vee^2 \bR^N \oplus
\vee^2S\otimes \La^2\bR^N\, .
\end{eqnarray*}
In particular, this implies that the multiplicities $\mu_+(k,N)$ and
$\mu_-(k,N)$  of the module $\La^kV$ in $\vee^2W_1$ and $\La^2W_1$,
respectively, are given by
\begin{eqnarray*}\mu_+(k,N) &=& \mu_+(k)\frac{N(N+1)}{2}
+ \mu_-(k)\frac{N(N-1)}{2}\,,\\
\mu_-(k,N) &=& \mu_-(k)\frac{N(N+1)}{2} + \mu_+(k)\frac{N(N-1)}{2}\, ,
\end{eqnarray*}
where $\mu_+(k)$ and $\mu_-(k)$ are the multiplicities of $\La^kV$ in
$\vee^2S$ and $\La^2S$, respectively.
The vector space of N-extended polyvector Poincar{\'e} $\e$-algebra structures
with $W_0 = \La^kV$ is identified with the space $\Bil_\e^k(W_1)^{\so(V)}$
of invariant $\La^kV$-valued bilinear forms on $W_1$. Its dimension is
given by
\[ \dim \Bil_\e^k(W_1)^{\so(V)} = \mu_\e(k,N) \dim \cc (\La^kV) \, ,\]
where   the Schur algebra $\cc (\La^kV) = \bR, \bR \oplus \bR$ or $\bC$, see
the proof of Theorem \ref{mainThm}.
Any element $\G_{\e} \in \Bil_\e^k(W_1)^{\so(V)}$ can be represented
as
\begin{eqnarray*} \G_+ &=& \sum_i \G^k_{\b_+^i}\otimes b^i_+ +
 \sum_j \G^k_{\b_-^j}\otimes b_-^j\,,\\
\G_- &=& \sum_i \G^k_{\b_-^i}\otimes b^i_+ +
 \sum_j \G^k_{\b_+^j}\otimes b_-^j\, ,
\end{eqnarray*}
where $\b_\pm^i \in \Bil_\pm^k(S)^{\so(V)}$ and  $b_+^i$ and  $b_-^i$
are, respectively, symmetric and skewsymmetric bilinear forms on $\bR^N$.
We note also, that there exists a unique minimal (i.e.\ $W_0 = [W_1,W_1]$)
$N$-extended polyvector Poincar{\'e}
$\e$-algebra with $W_0 = \mu_\e (k,N)\La^kV$. The Lie (super)bracket
is given, up to a twist by an invertible
element of the Schur algebra of $W_0$,
by the projection onto the corresponding maximal isotypical
submodule of $\vee^2W_1$ or $\La^2W_1$, respectively.

Similarly, in the case when the spinor module
$S = S_+ \oplus S_-$ is reducible, we can reduce the
description of all $(N_+,N_-)$-extended polyvector Poincar{\'e} $\e$-algebras
$\gg = \so(V) + \La^kV + W_1$ such that $W_1 = N_+S_+ + N_-S_-$ to the chiral
cases $(N_+,N_-) = (1,0)$ or $(0,1)$  and to the isotropic case:
$(N_+,N_-) = (1,1)$ and $[S_+,S_+ ] = [S_-,S_-] = 0$.

Let $\beta$ be an admissible bilinear form on $S= S_+ \oplus S_-$
and $\Gamma_\beta^k
\in \Bil_\e^k(S)^{\so(V)}$ the corresponding admissible $\La^kV$-valued
bilinear form. Its restriction to $S_+$ (or $S_-$) defines a
$(1,0)$-extended (respectively, $(0,1)$-extended)
$k$-polyvector Poincar{\'e} $\e$-algebra if and only if $\iota (\Gamma_\beta^k)
= +1$. If $\iota (\Gamma_\beta^k)
= -1$ then we obtain an isotropic $(1,1)$-extended
$k$-polyvector Poincar{\'e} $\e$-algebra, i.e.\ $[S_+,S_+ ] = [S_-,S_-] = 0$.
The values of the invariants $\sigma (\Gamma_\beta^k) = \e$
and $\iota (\Gamma_\beta^k)$ can be read off Tables
\ref{tbl:beta}--\ref{tbl:sik3} in Appendix~\ref{app:admissForms}.

\medskip
\section*{Acknowledgments}

\noindent This work was generously supported by the `Schwerpunktprogramm
Stringtheorie' of the Deutsche For\-schungs\-ge\-mein\-schaft. Partial
support by the European Commission's  5th Framework Human Potential
Programmes under contracts HPRN-CT-2000-00131 (Quantum  Spacetime) and
HPRN-CT-2002-00325 (EUCLID) is also acknowledged. A.V.P. is supported in
part by the Belgian Federal Office for Scientific, Technical and Cultural
Affairs through the Inter-University Attraction Pole P5/27. V.C, C.D. and
A.V.P. are grateful to the University of Hull for hospitality during a
visit in which the basis of this work was discussed.

\renewcommand{\thetable}{\thesection\arabic{table}}
\catcode`@=11
\@addtoreset{table}{section}
\catcode`@=12

\hfill
\appendix
\section{Admissible $\boldsymbol{\La^kV}$-valued bilinear forms on
$\boldsymbol{S}$} \label{app:admissForms}
In Table \ref{tabA0} we give the numbers $(N^+ (s,n),N^-(s,n))$
of independent symmetric
($+$) and skewsymmetric ($-$) invariant  bilinear forms on $S$, i.e.\
$N^\pm (s,n) = \dim \Bil_\pm (S)^{\so(V)}$. They were computed in \cite{AC}
and are periodic with period $8$ in the signature $s = p-q$ and in the
dimension $n = p+q$ of $V=\bR^{p,q}$. The entry $N(s)= N^+(s,n)+N^-(s,n)$
is the total number of invariant bilinear forms and
$\mu(k)$ is the multiplicity  of the irreducible submodule $\La^kV$ in
$S\ot S$. For $k\neq n/2$, it does not depend on $k$, i.e.\ $\mu (k) =
\mu (0)$. In the case  $n=2m$ the multiplicities $\mu(m)$ with a $*$ indicate
that the module  $\La^mV$ is reducible as $\La^m_+V + \La^m_- V $.
\begin{table}[h]
\begin{center}
\vskip 1 true cm
\begin{tabular}
{|c ||l    |l     |l      |l       |l      |l      |l     |l || c| c| c|}
\hline
$s\backslash n$  &0  &1 &2 &3 &4 &5 &6 &7 & $N(s)$ &$\mu(k)=\mu(0)$&$\mu(m)$
   \\  \hline
0 &  2,0 && 1,1 && 0,2 && 1,1 && 2 & 2&1*\\  \hline
1 && 3,1 && 1,3 && 1,3 && 3,1 & 4& 4&\\   \hline
2 & 6,2 && 4,4 && 2,6 && 4,4 && 8 & 8&4\\   \hline
3 && 3,1 && 1,3 && 1,3 && 3,1 & 4& 4&\\   \hline
4  & 6,2 && 4,4 && 2,6 && 4,4 && 8& 8&4*\\   \hline
5 && 3,1 && 1,3 && 1,3 && 3,1 & 4& 4&\\   \hline
6 &  2,0 && 1,1 && 0,2 && 1,1 && 2& 2&1\\  \hline
7 && 1,0 && 0,1 && 0,1 && 1,0 &1 &1&\\  \hline
%\hline
\end{tabular}\end{center}
\caption{\it The numbers $(N^+ (s,n),N^-(s,n))$ of independent symmetric and
skewsymmetric bilinear forms on $S$} \label{tabA0}
\end{table}
\newpage
\noindent
{}From Table \ref{tabA0}, we note the following symmetries:

\noindent
a) Modulo 8-symmetry
$$
 N^\pm(s+8a, n+8b) = N^\pm(s,n) \quad,\quad a,b\in\bZ
$$
b) Reflection with respect to the horizontal line $s=3$
$$ N^\pm(-s+6,n) = N^\pm(s,n)
$$
and reflection with respect to the vertical line $n=0$
$$N^\pm(s,n) = N^\pm(s,-n) \, .$$
c) The reflection with respect to the vertical line $n=2$ is a
mirror symmetry, i.e.\ it interchanges $N^+$ and $N^-$:
$$  N^\pm(s,n) = N^\mp(s,-n+4) \, .$$
It is the same as the mirror symmetry with respect to $n=6$.
We note that the composition of reflections in $n=0$ and $n=2$
gives the translation $n \mapsto n+4$:
$$ N^\pm(s,n) = N^\mp(s,n+4)\, .$$
More generally, we consider the dimension
\[ \tilde{N}_k^\pm (p,q) := N_k^\pm(s,n) := \dim\Bil^k_\pm (S)^{\so(V)}\]
of the vector space of invariant $\La^kV$-valued bilinear
forms on $S$. By Theorem \ref{mainThm}, the sum
$N_k(s,n) = N_k^+(s,n) + N_k^-(s,n) = N(s,n)$ does not depend on $k$.
{}From Table \ref{tabA0}, we see that it depends only on the signature $s$.
As a corollary of Proposition \ref{mainProp}, the numbers
$\tilde{N}_k^\pm (p,q)$ are periodic modulo $4$ in $k$:
\[ \tilde{N}_k^\pm (p,q) = \tilde{N}_{k+4}^\pm (p,q)\, .\]
Moreover, we have the following periodicities in $(p,q)$:
\[ \tilde{N}_k^\pm (p,q)= \tilde{N}_k^\pm (p+8,q) =  \tilde{N}_k^\pm (p,q+8)
=  \tilde{N}_k^\pm (p+4,q+4) \, .\]
In fact, it was proven in \cite{AC} that, for any
given symmetry $\s_0$, type $\t_0$ and isotropy $\iota_0$ (if defined),
the number of bilinear forms $\beta$ with $\s (\beta ) = \s_0$,
$\t (\beta ) = \t_0$ and $\iota (\beta ) = \iota_0$
in a basis $(\beta_i)$ of $\Bil (S)^{\so(V)}$ consisting of admissible forms
is $(8,0)$-, $(0,8)$- and $(4,4)$-periodic in $(p,q)$. By
Proposition \ref{mainProp}, this implies that for any
given symmetry $\s_0'$ and isotropy $\iota_0'$ (if defined),
the number of $\La^kV$-valued bilinear forms $\G_{\beta_i}^k$ with
$\s (\G_{\beta_i}^k) = \s_0'$, and $\iota (\G_{\beta_i}^k) = \iota_0'$
is $(8,0)$-, $(0,8)$- and $(4,4)$-periodic in $(p,q)$.

Finally, we have the following  shift formula
$$  N_k^\pm(s,n+2k) = N_0^\pm(s,n) :=  N^\pm(s,n)\, ,$$
which we can write also as
\[ \tilde{N}_k^\pm (p+k,q+k) = \tilde{N}_0^\pm(p,q)\, .\]
This shift formula follows from the tables below.

In Table \ref{tbl:beta} we describe a basis  of $\Bil(S)^{\so(V)}$,
which consists of admissible forms and indicate the values of
the three invariants $\s$, $\t$ and $\iota$. In the three following tables
we give the invariants $\s$ and $\iota$ for the corresponding
bases of $\Bil^k(S)^{\so(V)}$, $k = 1, 2, 3$ modulo 4,
denoted for simplicity by the same symbols.
Due to the above periodicity properties, we can calculate, from these tables,
the values of the invariants for the corresponding bases of
$\Bil^k(S)^{\so(V)}$ for all $k \in \bN$ and $V=\bR^{p,q}$.

For any  $V=\bR^{p,q}$ an explicit basis of $\Bil^k(S)^{\so(V)}$ consisting
of admissible forms was constructed for $k=0$ and $k=1$, in terms
appropriate models of the spinor module in \cite{AC}.
It was proven there that any admissible
$V$-valued bilinear form on $S$ is of the form $\Gamma_\beta^1$,
where $\beta$ is a linear combination of admissible
scalar-valued forms. By Theorem \ref{mainThm}
and Proposition \ref{mainProp}, this result extends to $k>1$, namely
any admissible $\La^kV$-valued bilinear form is of the
form $\Gamma_\beta^k$, where $\beta$ is a linear combination of admissible
scalar-valued forms. This provides a basis
of the vector space ${\rm Bil}^k (S)^{\so(V)} \cong {\rm Bil} (S)^{\so(V)}$
consisting of admissible forms. The dimension of this space is equal to the
dimension of the Schur algebra $\cc (S)$, which depends only on the signature
$s = p-q$ modulo $8$, see \cite{AC}.

In the tables below we use the notation of \cite{AC}. We use the fact that
any pseudo-Euclidean vector space can be written as $V = V_1 \oplus V_2$,
where $V_1 = \bR^{r,r}$ and $V_2 = \bR^{0,k}$ or  $V_2 = \bR^{k,0}$.
Then  \cite{LM}
\[ \Cl (V) \cong \Cl (V_1) \;\wh\ot\; \Cl (V_2)\, ,\]
where $\wh\ot$ denotes the $\bZ/2\bZ$-graded tensor product.
Let $S_1$ be the spinor module of Spin($V_1$) and
$S_2$ the spinor module of Spin($V_2$). Then we always have that
$S_1 = S_1^+ + S_1^-$ is a $\bZ/2\bZ$-graded module of the
$\bZ/2\bZ$-graded algebra $\Cl (V_1)$. The spinor module $S$
of Spin($V$) can be described in terms of $S_1$ and $S_2$ as follows, see
Proposition 2.3. of \cite{AC}.
Consider first the case when $S_2 = S_2^+ + S_2^-$ is reducible. In this
case $S_2$ is a $\bZ/2\bZ$-graded $\Cl (V_2)$-module and the spinor
module $S = S_1 \;\wh\ot\; S_2$ of Spin($V$)
is obtained as the  $\bZ/2\bZ$-graded tensor product of
the modules $S_1$ and $S_2$. It is again $\bZ/2\bZ$-graded with even part
$S_+ = S_1^+\otimes S_2^+ + S_1^-\otimes S_2^-$ and odd part
$S_- = S_1^+\otimes S_2^- + S_1^-\otimes S_2^+$.
If $S_2$ is an irreducible Spin($V$)-module, then the spinor module
of Spin($V$) is given by
\[ S = S_1 \;\wh\ot\; S_2 = S_1^+ \otimes S_2
+ S_1^- \otimes S_2\]
with the action
\[ (a \;\wh\ot\; b) \cdot (s_1^\pm \otimes s_2) = (-1)^{{\rm deg}(s_1^\pm)
{\rm deg}(b)}
as_1^\pm \otimes bs_2  \, ,\]
where $a \in \Cl (V_1)$, $b\in  \Cl (V_1)$, $s_1^\pm \in S_1^\pm$,
$s_2 \in S_2$, ${\rm deg}(s_1^+) = 0$ and ${\rm deg}(s_1^-) = 1$.
In this case $S$ is an irreducible Spin($V$)-module.

As discussed in Section \ref{complex}, for the case of split
signature, $V=\bR^{r,r}$, there exist two independent admissible
bilinear forms $f$ and  $f_E=f(E\cdot\,,\,\cdot)$ on the spinor module,
where $E|_{S_\pm}=\pm {\rm Id}$. Their invariants are given in Table
\ref{tbl:beta}. If $V$ is positive or negative definite, then there exists
a unique (up to scale) ${\rm Pin}(V)$-invariant scalar product
$g$ on the spinor module $S$ and any admissible, hence invariant,
bilinear form on $S$ is of the form $g_A = g(A\cdot ,\cdot )$,
where $A$ is an admissible element of the Schur algebra $\cc (S)$.
The admissibility of $A$ means that $A$ is either symmetric or skewsymmetric
with respect to $g$, that it either commutes or anticommutes with the
Clifford multiplication $\g_v$ and that it preserves or interchanges
$S_+$ and $S_-$ if they exist \cite{AC}.

\begin{table}\begin{center}
\vskip 1 true cm
\begin{tabular}
{|c   |c        |c   | } \hline
 $s$ & $\cc (S)$ &  basis of $\cc (S)$\\ \hline
0 & $2\bR$ & ${\rm Id}, E$  \\
1 &$\bR(2)$  & ${\rm Id}, E, I, EI$\\
2& $\bC(2)$ & ${\rm Id}, I, J, K, E, EI, EJ, EK$\\
3& $\bH$ & ${\rm Id}, I, J, K$\\
4& $2\bH$ & ${\rm Id}, I, J, K,E, EI, EJ, EK$\\
5& $\bH$ & ${\rm Id}, I, J, K$\\
6 & $\bC$ & ${\rm Id}, I$\\
7& $\bR$& ${\rm Id}$\\
\hline
\end{tabular}
\end{center}
\caption{\it Standard bases for the Schur algebras $\cc (S)$} \label{schur}
\end{table}

Returning to the general case $V = \bR^{p,q} = V_1 \oplus V_2$, $S = S_1
\;\wh\ot\; S_2$, as above, the admissible bilinear forms on $S$ can be
described as follows. Let $(g_{A_i})$ be a basis of ${\rm Bil}
(S_2)^{\so(V_2)}$ consisting of admissible elements. Inspection of Table
\ref{tbl:beta} shows that for any $g_{A_i}$ there exists a unique element
$\phi_i \in \{ f, f_E\}$ which satisfies the condition $\t (\phi_i ) =
\iota (g_{A_i}) \t (g_{A_i})$. In virtue of Proposition 3.4 of \cite{AC},
the tensor products $\phi_i \otimes g_{A_i}$ provide a basis of ${\rm
Bil} (S)^{\so(V)}$ consisting of admissible elements. The corresponding
basis $\G^k_{\phi_i \otimes g_{A_i}}$ of ${\rm Bil}^k (S)^{\so(V)}$ and
its invariants are tabulated below. For simplicity the symbols $\G^k$ and
$\otimes$ are omitted.

We use the following bases for the Schur algebra
$\cc (S)$ (see Table \ref{schur}). If the signature $s = 0, 1, 2$ or $4$,
then  $S = S_+ \oplus S_-$, and we put $E = {\rm diag} ({\rm Id}_{S_+},
{\rm Id}_{S_-})$. In the cases $s = 1$ and $s = 6$ we denote by $I$
the standard complex structure in $\cc (S) = \bR (2)$ and $\cc (S) = \bC$,
respectively. In fact, in the case $s = 1$, $S = \bC^N = \bR^N \oplus
\bR^N$ has a $\Cl (V)$-invariant complex structure.
In the cases $s = 2, 3, 4$ and $5$, we denote by
$I, J, K = IJ\in \cc (S)$ the canonical $\Cl^0 (V)$-invariant
hypercomplex structure of $S$ (3 anticommuting complex structures).
In fact, in the case $s = 4$, $S = \bH^{N/2} =  \bH^{N/4} \oplus  \bH^{N/4}$
has a $\Cl (V)$-invariant hypercomplex structure.

%%
%% k=0 table
%%

\begin{table}
\begin{center}\footnotesize
\caption{\small Admissible bilinear forms $\beta$ on $S$ and their
invariants $(\sigma,\tau,\iota)$ \label{tbl:beta}}
\begin{tabular}
{|c |ll             |ll                  |ll               |ll | } \hline
 p$\backslash$q  &&0  &&1                 &&2 &&3
   \\  \hline
0  &$f$ & $+-+$     &$g$&$++$        &$fg$  &$--$    &$f_Eg$ &$-+$\\[-1pt]
   &$f_E$&$+++$     &&                  &$f_Eg_I$&$++$  &$f_Eg_I$&$++$\\[-1pt]
   &&               &&                  &&              &$f_Eg_J$&$--$\\[-1pt]
   &&               &&                  &&              &$f_Eg_K$&$--$\\
\hline
1    &$g$ & $+-+$    &$f$  &$---$        &$g$  &$--$    &$f_Eg$ &$-+$\\[-1pt]
     &$g_E$&$+++$    &$f_E$&$++-$        &&              &$fg_I$ &$--$\\[-1pt]
 &$g_{I}$&$---$    &&                  &&              &&\\[-1pt]
&$g_{IE}$&$++-$    &&                  &&              &&\\
\hline
2   &$g$ &  $+-+$  &$f_Eg$      &$++-$   &$f$  &$--+$    &$g$ &$-+$\\[-1pt]
 &$g_{I}$&$--+$  &$f_Eg_{I}$&$-++$   &$f_E$&$-++$    &&\\[-1pt]
 &$g_{J}$&$---$  &$fg_E$      &$---$   &&              &&\\[-1pt]
 &$g_{K}$&$---$  &$fg_{IE}$ &$--+$   &&              &&\\[-1pt]
     &$g_E$&$+++$  &&                    &&              &&\\[-1pt]
&$g_{IE}$&$-++$  &&                    &&              &&\\[-1pt]
&$g_{JE}$&$++-$  &&                    &&              &&\\[-1pt]
&$g_{KE}$&$++-$  &&                    &&              &&\\
\hline
3   &$g$ &  $+-$ &$f_Eg$      &$++-$ &$fg$      &$--+$  &$f$  &$+--$\\[-1pt]
 &$g_{I}$&$--$ &$f_Eg_{I}$&$-+-$ &$fg_{I}$ &$+--$ &$f_E$&$-+-$\\[-1pt]
 &$g_{J}$&$--$  &$f_Eg_{J}$&$-++$  &$f_Eg_E$     &$-++$   &&\\[-1pt]
 &$g_{K}$&$--$  &$f_Eg_{K}$&$-++$  &$f_Eg_{IE}$&$-+-$   &&\\[-1pt]
 &&               &$fg_E$      &$---$   &&              &&\\[-1pt]
 &&               &$fg_{IE}$ &$+--$   &&              &&\\[-1pt]
 &&               &$fg_{JE}$ &$--+$   &&              &&\\[-1pt]
 &&               &$fg_{KE}$ &$--+$   &&              &&\\
\hline
4  &$g$  &$+-+$   &$f_Eg$  &$++$  &$fg$  &$--+$   &$f_Eg$ &$-+-$\\[-1pt]
&$g_{I}$&$--+$  &$f_Eg_{I}$ &$-+$  &$fg_{I}$ &$+-+$&$f_Eg_{I}$&$+++$
\\[-1pt]
&$g_{J}$&$--+$ &$f_Eg_{J}$&$-+$ &$fg_{J}$&$+--$ &$fg_E$  &$+--$\\[-1pt]
&$g_{K}$&$--+$ &$f_Eg_{K}$&$-+$ &$fg_{K}$&$+--$ &$fg_{IE}$&$+-+$
\\[-1pt]
&$g_E  $&$+++$ &&                  &$f_Eg_E  $&$-++$  &&\\[-1pt]
&$g_{IE}$&$-++$ &&               &$f_Eg_{IE}$&$+++$  &&\\[-1pt]
&$g_{JE}$&$-++$ &&               &$f_Eg_{JE}$&$-+-$  &&\\[-1pt]
&$g_{KE}$&$-++$ &&               &$f_Eg_{KE}$&$-+-$  &&\\  \hline
5  &$g$&$+-$ &$f_Eg$      &$++-$ &$fg$      &$--$  &$f_Eg$     &$-+-$\\[-1pt]
 &$g_I$&$--$ &$f_Eg_{I}$&$-+-$ &$fg_{I}$&$+-$ &$f_Eg_{I}$&$++-$\\[-1pt]
 &$g_J$&$-+$ &$f_Eg_{J}$&$-+-$ &$fg_{J}$&$+-$ &$f_Eg_{J}$&$+++$\\[-1pt]
 &$g_K$&$-+$ &$f_Eg_{K}$&$-+-$ &$fg_{K}$&$+-$ &$f_Eg_{K}$&$+++$\\[-1pt]
&&          &$fg_E$      &$---$ &&                &$fg_E$ &$+--$\\[-1pt]
&&          &$fg_{IE}$ &$+--$ &&                &$fg_{IE}$ &$---$\\[-1pt]
&&          &$fg_{JE}$ &$+--$ &&                &$fg_{JE}$ &$+-+$\\[-1pt]
  &&          &$fg_{KE}$ &$+--$  &&                &$fg_{KE}$ &$+-+$\\
\hline
6  &$g$&$+-$   &$f_Eg$  &$++$    &$fg$      &$--+$  &$f_Eg$ &$-+$\\[-1pt]
&$g_I$&$-+$   &$f_Eg_I$&$-+$    &$fg_{I}$&$+-+$  &$f_Eg_{I}$&$++$\\[-1pt]
&&            &$fg$    &$+-$    &$fg_{J}$&$+-+$  &$f_Eg_{J}$&$++$\\[-1pt]
&&            &$fg_I$  &$+-$    &$fg_{K}$&$+-+$  &$f_Eg_{K}$&$++$\\[-1pt]
  &&            &&              &$f_Eg_E  $&$-++$  &&\\[-1pt]
  &&            &&              &$f_Eg_{IE}$&$+++$  &&\\[-1pt]
  &&            &&             &$f_Eg_{JE}$&$+++$  &&\\[-1pt]
  &&
  &&             &$f_Eg_{KE}$&$+++$  &&\\
\hline
7 &$g$&$+-$ &$f_Eg$&$++$   &$fg$    &$--$       &$f_Eg$    &$-+-$\\[-1pt]
   &&         &$fg_I$&$+-$   &$fg_I$  &$+-$     &$f_Eg_{I}$&$++-$\\[-1pt]
   &&         &&             &$f_Eg$  &$++$     &$f_Eg_{J}$&$++-$\\[-1pt]
   &&         &&             &$f_Eg_I$&$++$     &$f_Eg_{K}$&$++-$\\[-1pt]
   &&         &&                  &&             &$fg_E$     &$+--$\\[-1pt]
   &&         &&                  &&             &$fg_{IE}$&$---$\\[-1pt]
   &&         &&                  &&             &$fg_{JE}$&$---$\\[-1pt]
   &&         &&                  &&             &$fg_{KE}$&$---$\\
\hline
\end{tabular}  \end{center} \end{table}

%%
%% k=1 table
%%
\begin{table}\begin{center}\footnotesize
\caption{\small $\La^kV$-valued admissible bilinear forms on $S$ and
their invariants $(\sigma,\iota)$ for $k {\equiv} 1 (4)$
\label{tbl:sik1}}

\begin{tabular}
{|c |ll             |ll                  |ll               |ll | }
\hline
p$\backslash$q  &&0  &&1                 &&2              &&3
   \\  \hline
0  &$f$ & $--$     &$g$&$+$        &$fg$  &$+$    &$f_Eg$ &$-$\\[-1pt]
   &$f_E$&$+-$     &&                  &$f_Eg_I$&$+$  &$f_Eg_I$&$+$\\[-1pt]
   &&               &&                  &&              &$f_Eg_J$&$+$\\[-1pt]
   &&               &&                  &&              &$f_Eg_K$&$+$\\
\hline
1    &$g$ & $--$    &$f$  &$++$        &$g$  &$+$    &$f_Eg$ &$-$\\[-1pt]
     &$g_E$&$+-$    &$f_E$&$++$        &&              &$fg_I$ &$+$\\[-1pt]
 &$g_{I}$&$++$    &&                  &&              &&\\[-1pt]
&$g_{IE}$&$++$    &&                  &&              &&\\
\hline
2   &$g$ &  $--$  &$f_Eg$      &$++$   &$f$  &$+-$    &$g$ &$-$\\[-1pt]
 &$g_{I}$&$+-$  &$f_Eg_{I}$&$--$   &$f_E$&$--$    &&\\[-1pt]
 &$g_{J}$&$++$  &$fg_E$      &$++$   &&              &&\\[-1pt]
 &$g_{K}$&$++$  &$fg_{IE}$ &$+-$   &&              &&\\[-1pt]
     &$g_E$&$+-$  &&                    &&              &&\\[-1pt]
&$g_{IE}$&$--$  &&                    &&              &&\\[-1pt]
&$g_{JE}$&$++$  &&                    &&              &&\\[-1pt]
&$g_{KE}$&$++$  &&                    &&              &&\\
\hline
3   &$g$ &  $-$ &$f_Eg$     &$++$  &$fg$       &$+-$   &$f$  &$-+$\\[-1pt]
 &$g_{I}$&$+$ &$f_Eg_{I}$&$-+$ &$fg_{I}$ &$-+$   &$f_E$&$-+$\\[-1pt]
 &$g_{J}$&$+$  &$f_Eg_{J}$&$--$  &$f_Eg_E$     &$--$   &&\\[-1pt]
 &$g_{K}$&$+$  &$f_Eg_{K}$&$--$  &$f_Eg_{IE}$&$-+$   &&\\[-1pt]
 &&               &$fg_E$      &$++$   &&              &&\\[-1pt]
 &&               &$fg_{IE}$ &$-+$   &&              &&\\[-1pt]
 &&               &$fg_{JE}$ &$+-$   &&              &&\\[-1pt]
 &&               &$fg_{KE}$ &$+-$   &&              &&\\
\hline
4  &$g$   &$--$ &$f_Eg$  &$+$  &$fg$  &$+-$     &$f_Eg$  &$-+$\\[-1pt]
&$g_{I}$&$+-$ &$f_Eg_{I}$&$-$  &$fg_{I}$&$--$ &$f_Eg_{I}$&$+-$
\\[-1pt]
&$g_{J}$&$+-$ &$f_Eg_{J}$&$-$  &$fg_{J}$&$-+$ &$fg_E$&$-+$\\[-1pt]
&$g_{K}$&$+-$ &$f_Eg_{K}$&$-$  &$fg_{K}$&$-+$ &$fg_{IE} $&$--$
\\[-1pt]
&$g_E  $&$+-$ &&                  &$f_Eg_E  $&$--$  &&\\[-1pt]
&$g_{IE}$&$--$ &&               &$f_Eg_{IE}$&$+-$  &&\\[-1pt]
&$g_{JE}$&$--$ &&               &$f_Eg_{JE}$&$-+$  &&\\[-1pt]
&$g_{KE}$&$--$ &&               &$f_Eg_{KE}$&$-+$  &&\\
\hline
5  &$g$&$-$ &$f_Eg$      &$++$ &$fg$      &$+$  &$f_Eg$      &$-+$\\[-1pt]
 &$g_I$&$+$ &$f_Eg_{I}$&$-+$ &$fg_{I}$&$-$  &$f_Eg_{I}$&$++$\\[-1pt]
 &$g_J$&$-$ &$f_Eg_{J}$&$-+$ &$fg_{J}$&$-$  &$f_Eg_{J}$&$+-$\\[-1pt]
 &$g_K$&$-$ &$f_Eg_{K}$&$-+$ &$fg_{K}$&$-$  &$f_Eg_{K}$&$+-$\\[-1pt]
 &&          &$fg_E$      &$++$ &&                &$fg_E$      &$-+$\\[-1pt]
 &&          &$fg_{IE}$ &$-+$ &&                &$fg_{IE}$ &$++$\\[-1pt]
 &&          &$fg_{JE}$ &$-+$ &&                &$fg_{JE}$ &$--$\\[-1pt]
 &&          &$fg_{KE}$ &$-+$ &&                &$fg_{KE}$ &$--$\\
\hline
6  &$g$&$-$   &$f_Eg$  &$+$    &$fg$      &$+-$  &$f_Eg$      &$-$\\[-1pt]
 &$g_I$&$-$   &$f_Eg_I$&$-$    &$fg_{I}$&$--$  &$f_Eg_{I}$&$+$\\[-1pt]
 &&            &$fg$    &$-$    &$fg_{J}$&$--$  &$f_Eg_{J}$&$+$\\[-1pt]
 &&            &$fg_I$  &$-$    &$fg_{K}$&$--$  &$f_Eg_{K}$&$+$\\[-1pt]
 &&            &&                &$f_Eg_E  $&$--$  &&\\[-1pt]
 &&            &&             &$f_Eg_{IE}$&$+-$  &&\\[-1pt]
 &&            &&             &$f_Eg_{JE}$&$+-$  &&\\[-1pt]
 &&            &&             &$f_Eg_{KE}$&$+-$  &&\\
\hline
7 &$g$&$-$  &$f_Eg$&$+$   &$fg$    &$+$       &$f_Eg$    &$-+$\\[-1pt]
  &&          &$fg_I$&$-$   &$fg_I$  &$-$     &$f_Eg_{I}$&$++$\\[-1pt]
  &&          &&             &$f_Eg$  &$+$     &$f_Eg_{J}$&$++$\\[-1pt]
  &&          &&             &$f_Eg_I$&$+$     &$f_Eg_{K}$&$++$\\[-1pt]
  &&          &&                  &&             &$fg_E$     &$-+$\\[-1pt]
  &&          &&                  &&             &$fg_{IE}$&$++$\\[-1pt]
  &&          &&                  &&             &$fg_{JE}$&$++$\\[-1pt]
  &&          &&                  &&             &$fg_{KE}$&$++$\\
\hline
\end{tabular}
\end{center}
\end{table}
%%
%% k=2 table
%%
\begin{table}\begin{center}\footnotesize
\caption{\small $\La^kV$-valued admissible bilinear forms on $S$ and
their invariants $(\sigma,\iota)$ for $k {\equiv} 2 (4)$
\label{tbl:sik2}}
\begin{tabular}
{|c |ll             |ll                  |ll               |ll | }
\hline
p$\backslash$q  &&0  &&1                 &&2              &&3
   \\  \hline
0  &$f$ & $-+$    &$g$&$-$        &$fg$  &$+$    &$f_Eg$ &$+$\\[-1pt]
   &$f_E$&$-+$    &&                  &$f_Eg_I$&$-$  &$f_Eg_I$&$-$\\[-1pt]
   &&              &&                  &&              &$f_Eg_J$&$+$\\[-1pt]
   &&              &&                  &&              &$f_Eg_K$&$+$\\
\hline
1    &$g$ & $-+$   &$f$  &$+-$        &$g$  &$+$    &$f_Eg$ &$+$\\[-1pt]
     &$g_E$&$-+$   &$f_E$&$--$        &&              &$fg_I$ &$+$\\[-1pt]
 &$g_{I}$&$+-$    &&                  &&              &&\\[-1pt]
&$g_{IE}$&$--$    &&                  &&              &&\\
\hline
2   &$g$ &  $-+$  &$f_Eg$      &$--$   &$f$  &$++$    &$g$ &$+$\\[-1pt]
 &$g_{I}$&$++$  &$f_Eg_{I}$&$++$   &$f_E$&$++$    &&\\[-1pt]
 &$g_{J}$&$+-$  &$fg_E$      &$+-$   &&              &&\\[-1pt]
 &$g_{K}$&$+-$  &$fg_{IE}$ &$++$   &&              &&\\[-1pt]
     &$g_E$&$-+$  &&                    &&              &&\\[-1pt]
&$g_{IE}$&$++$  &&                    &&              &&\\[-1pt]
&$g_{JE}$&$--$  &&                    &&              &&\\[-1pt]
&$g_{KE}$&$--$  &&                    &&              &&\\
\hline
3   &$g$ &  $-$ &$f_Eg$      &$--$  &$fg$       &$++$ &$f$  &$--$\\[-1pt]
 &$g_{I}$&$+$ &$f_Eg_{I}$&$+-$  &$fg_{I}$ &$--$ &$f_E$&$+-$\\[-1pt]
 &$g_{J}$&$+$  &$f_Eg_{J}$&$++$  &$f_Eg_E$     &$++$   &&\\[-1pt]
 &$g_{K}$&$+$  &$f_Eg_{K}$&$++$  &$f_Eg_{IE}$&$+-$   &&\\[-1pt]
 &&               &$fg_E$      &$+-$   &&              &&\\[-1pt]
 &&               &$fg_{IE}$ &$--$   &&              &&\\[-1pt]
 &&               &$fg_{JE}$ &$++$   &&              &&\\[-1pt]
 &&               &$fg_{KE}$ &$++$   &&              &&\\
\hline
4  &$g$  &$-+$ &$f_Eg$ &$-$  &$fg$  &$++$  &$f_Eg$  &$+-$ \\[-1pt]
&$g_{I}$&$++$ &$f_Eg_{I}$&$+$  &$fg_{I}$&$-+$ &$f_Eg_{I}$&$-+$
\\[-1pt]
&$g_{J}$&$++$ &$f_Eg_{J}$&$+$  &$fg_{J}$&$--$ &$fg_E$      &$--$
\\[-1pt]
&$g_{K}$&$++$ &$f_Eg_{K}$&$+$  &$fg_{K}$&$--$ &$fg_{IE} $&$-+$
\\[-1pt]
&$g_E  $&$-+$ &&                  &$f_Eg_E  $&$++$  &&\\[-1pt]
&$g_{IE}$&$++$ &&               &$f_Eg_{IE}$&$-+$  &&\\[-1pt]
&$g_{JE}$&$++$ &&               &$f_Eg_{JE}$&$+-$  &&\\[-1pt]
&$g_{KE}$&$++$ &&               &$f_Eg_{KE}$&$+-$  &&\\
\hline
5  &$g$&$-$ &$f_Eg$      &$--$ &$fg$      &$+$  &$f_Eg$      &$+-$\\[-1pt]
 &$g_I$&$+$ &$f_Eg_{I}$&$+-$ &$fg_{I}$&$-$  &$f_Eg_{I}$&$--$\\[-1pt]
 &$g_J$&$+$ &$f_Eg_{J}$&$+-$ &$fg_{J}$&$-$  &$f_Eg_{J}$&$-+$\\[-1pt]
 &$g_K$&$+$ &$f_Eg_{K}$&$+-$ &$fg_{K}$&$-$  &$f_Eg_{K}$&$-+$\\[-1pt]
 &&          &$fg_E$      &$+-$ &&                &$fg_E$      &$--$\\[-1pt]
 &&          &$fg_{IE}$ &$--$ &&                &$fg_{IE}$ &$+-$\\[-1pt]
 &&          &$fg_{JE}$ &$--$ &&                &$fg_{JE}$ &$-+$\\[-1pt]
 &&          &$fg_{KE}$ &$--$ &&                &$fg_{KE}$ &$-+$\\
\hline
6  &$g$&$-$   &$f_Eg$  &$-$    &$fg$      &$++$  &$f_Eg$      &$+$\\[-1pt]
 &$g_I$&$+$   &$f_Eg_I$&$+$    &$fg_{I}$&$-+$  &$f_Eg_{I}$&$-$\\[-1pt]
 &&            &$fg$    &$-$    &$fg_{J}$&$-+$  &$f_Eg_{J}$&$-$\\[-1pt]
 &&            &$fg_I$  &$-$    &$fg_{K}$&$-+$  &$f_Eg_{K}$&$-$\\[-1pt]
 &&            &&                &$f_Eg_E  $&$++$  &&\\[-1pt]
 &&            &&             &$f_Eg_{IE}$&$-+$  &&\\[-1pt]
 &&            &&             &$f_Eg_{JE}$&$-+$  &&\\[-1pt]
 &&            &&             &$f_Eg_{KE}$&$-+$  &&\\
\hline
7 &$g$&$-$  &$f_Eg$&$-$   &$fg$    &$+$       &$f_Eg$    &$+-$\\[-1pt]
  &&          &$fg_I$&$-$   &$fg_I$  &$-$     &$f_Eg_{I}$&$--$\\[-1pt]
  &&          &&             &$f_Eg$  &$-$     &$f_Eg_{J}$&$--$\\[-1pt]
  &&          &&             &$f_Eg_I$&$-$     &$f_Eg_{K}$&$--$\\[-1pt]
  &&          &&                  &&             &$fg_E$     &$--$\\[-1pt]
  &&          &&                  &&             &$fg_{IE}$&$+-$\\[-1pt]
  &&          &&                  &&             &$fg_{JE}$&$+-$\\[-1pt]
  &&          &&                  &&             &$fg_{KE}$&$+-$\\
\hline
\end{tabular}
\end{center}
\end{table}
%%
%% k=3 table
%%
\begin{table}\begin{center}\footnotesize
 \caption{\small $\La^kV$-valued admissible bilinear forms on
$S$ and their invariants $(\sigma,\iota)$ for $k {\equiv} 3 (4)$
\label{tbl:sik3}}
\begin{tabular}
{|c |ll             |ll                  |ll               |ll | }
\hline
p$\backslash$q  &&0  &&1                 &&2              &&3
   \\  \hline
0  &$f$ & $+-$    &$g$&$-$        &$fg$  &$-$    &$f_Eg$ &$+$\\[-1pt]
   &$f_E$&$--$    &&                  &$f_Eg_I$&$-$  &$f_Eg_I$&$-$\\[-1pt]
   &&              &&                  &&              &$f_Eg_J$&$-$\\[-1pt]
   &&              &&                  &&              &$f_Eg_K$&$-$\\
\hline
1    &$g$ & $+-$   &$f$  &$-+$        &$g$  &$-$    &$f_Eg$ &$+$\\[-1pt]
     &$g_E$&$--$   &$f_E$&$-+$        &&              &$fg_I$ &$-$\\[-1pt]
 &$g_{I}$&$-+$   &&                  &&              &&\\[-1pt]
&$g_{IE}$&$-+$   &&                  &&              &&\\
\hline
2   &$g$ &  $+-$  &$f_Eg$      &$-+$   &$f$  &$--$   &$g$ &$+$\\[-1pt]
 &$g_{I}$&$--$  &$f_Eg_{I}$&$+-$   &$f_E$&$+-$   &&\\[-1pt]
 &$g_{J}$&$-+$  &$fg_E$      &$-+$   &&             &&\\[-1pt]
 &$g_{K}$&$-+$  &$fg_{IE}$ &$--$   &&             &&\\[-1pt]
     &$g_E$&$--$  &&                    &&             &&\\[-1pt]
&$g_{IE}$&$+-$  &&                    &&             &&\\[-1pt]
&$g_{JE}$&$-+$  &&                    &&             &&\\[-1pt]
&$g_{KE}$&$-+$  &&                    &&             &&\\
\hline
3   &$g$ &  $+$ &$f_Eg$    &$-+$  &$fg$        &$--$  &$f$ &$++$\\[-1pt]
&$g_{I}$&$-$ &$f_Eg_{I}$&$++$  &$fg_{I}$ &$++$  &$f_E$&$++$\\[-1pt]
 &$g_{J}$&$-$  &$f_Eg_{J}$&$+-$  &$f_Eg_E$   &$+-$   &&\\[-1pt]
&$g_{K}$&$-$  &$f_Eg_{K}$&$+-$  &$f_Eg_{IE}$&$++$   &&\\[-1pt]
&&  &$fg_E$      &$-+$   &&              &&\\[-1pt]  &&
 &$fg_{IE}$ &$++$   &&              &&\\[-1pt]  &&
&$fg_{JE}$ &$--$   &&              &&\\[-1pt]  &&
&$fg_{KE}$ &$--$   &&              &&\\ \hline
4  &$g$  &$+-$ &$f_Eg$  &$-$  &$fg$ &$--$  &$f_Eg$  &$++$\\[-1pt]
&$g_{I}$&$--$ &$f_Eg_{I}$&$+$  &$fg_{I}$&$+-$ &$f_Eg_{I}$&$--$
\\[-1pt]
&$g_{J}$&$--$ &$f_Eg_{J}$&$+$  &$fg_{J}$&$++$ &$fg_E$      &$++$
\\[-1pt]
&$g_{K}$&$--$ &$f_Eg_{K}$&$+$  &$fg_{K}$&$++$ &$fg_{IE} $&$+-$
\\[-1pt]
&$g_E  $&$--$ &&                  &$f_Eg_E  $&$+-$  &&\\[-1pt]
&$g_{IE}$&$+-$ &&               &$f_Eg_{IE}$&$--$  &&\\[-1pt]
&$g_{JE}$&$+-$ &&               &$f_Eg_{JE}$&$++$  &&\\[-1pt]
&$g_{KE}$&$+-$ &&               &$f_Eg_{KE}$&$++$  &&\\
\hline
5  &$g$&$+$ &$f_Eg$      &$-+$ &$fg$      &$-$  &$f_Eg$      &$++$\\[-1pt]
 &$g_I$&$-$ &$f_Eg_{I}$&$++$ &$fg_{I}$&$+$  &$f_Eg_{I}$&$-+$\\[-1pt]
 &$g_J$&$+$ &$f_Eg_{J}$&$++$ &$fg_{J}$&$+$  &$f_Eg_{J}$&$--$\\[-1pt]
 &$g_K$&$+$ &$f_Eg_{K}$&$++$ &$fg_{K}$&$+$  &$f_Eg_{K}$&$--$\\[-1pt]
 &&          &$fg_E$      &$-+$ &&                &$fg_E$      &$++$\\[-1pt]
 &&          &$fg_{IE}$ &$++$ &&                &$fg_{IE}$ &$-+$\\[-1pt]
 &&          &$fg_{JE}$ &$++$ &&                &$fg_{JE}$ &$+-$\\[-1pt]
 &&          &$fg_{KE}$ &$++$ &&                &$fg_{KE}$ &$+-$\\
\hline
6  &$g$&$+$   &$f_Eg$  &$-$    &$fg$      &$--$  &$f_Eg$      &$+$\\[-1pt]
 &$g_I$&$+$   &$f_Eg_I$&$+$    &$fg_{I}$&$+-$  &$f_Eg_{I}$&$-$\\[-1pt]
 &&            &$fg$    &$+$    &$fg_{J}$&$+-$  &$f_Eg_{J}$&$-$\\[-1pt]
 &&            &$fg_I$  &$+$    &$fg_{K}$&$+-$  &$f_Eg_{K}$&$-$\\[-1pt]
 &&            &&                &$f_Eg_E  $&$+-$  &&\\[-1pt]
 &&            &&             &$f_Eg_{IE}$&$--$  &&\\[-1pt]
 &&            &&             &$f_Eg_{JE}$&$--$  &&\\[-1pt]
 &&            &&             &$f_Eg_{KE}$&$--$  &&\\
\hline
7 &$g$&$+$  &$f_Eg$&$-$   &$fg$    &$-$       &$f_Eg$    &$++$\\[-1pt]
  &&          &$fg_I$&$+$   &$fg_I$  &$+$     &$f_Eg_{I}$&$-+$\\[-1pt]
  &&          &&             &$f_Eg$  &$-$     &$f_Eg_{J}$&$-+$\\[-1pt]
  &&          &&             &$f_Eg_I$&$-$     &$f_Eg_{K}$&$-+$\\[-1pt]
  &&          &&                  &&             &$fg_E$     &$++$\\[-1pt]
  &&          &&                  &&             &$fg_{IE}$&$-+$\\[-1pt]
  &&          &&                  &&             &$fg_{JE}$&$-+$\\[-1pt]
  &&          &&                  &&             &$fg_{KE}$&$-+$\\
\hline
\end{tabular}
\end{center}
\end{table}

\newpage
\section{Reformulation for physicists} \label{app:reformPhysicists}
In this appendix, we reformulate our results in a language that may be
more familiar to physicists. It is useful first to review some properties
of Clifford algebras, in particular those that concern the real Clifford
algebras.

\subsection{Complex and real Clifford algebras}\label{app:ClComReal}

We use here the terminology of Clifford algebras, spinors and gamma
matrices as used in physics. Results for the real case are dependent on
the signature. We remark that in the main text Clifford algebras have
been taken with a minus sign:
\begin{equation}
  \gamma ^a\gamma ^b+\gamma ^b\gamma ^a=-2\eta ^{ab}\,.
 \label{Clifford-}
\end{equation}
The signature $s$ has been introduced as $p-q$ modulo 8, where $p$ and
$q$ are the number of $+1$, respectively $-1$ eigenvalues of the metric
$\eta ^{ab}$. Thus,
\begin{eqnarray}
  p&=&\mbox{number of negative eigenvalues of }\left(\gamma ^a\gamma ^b+\gamma ^b\gamma ^a\right)
  \nonumber\\
  q&=&\mbox{number of positive eigenvalues of }\left(\gamma ^a\gamma ^b+\gamma ^b\gamma ^a\right)
  \nonumber\\
  s&=& p-q\ \mod 8\,,\qquad n=p+q\,.
 \label{defsneg}
\end{eqnarray}
This is important in order to interpret Table~\ref{TabCl}.

The bilinear form $\beta $ corresponds to the charge conjugation matrix
${\cal C}$, or for spinors $s$ and $t$, we have $\beta (s,t)=s^T{\cal
C}t$. If $v=e^a$, a basis vector, then the operation $\gamma (v)$ is the
gamma matrix $\gamma(v=e^a) = \gamma ^a$. The invariant $\sigma ({\cal
C})$ indicates the symmetry of ${\cal C}$, while $\tau ({\cal C})$
indicates the symmetry of ${\cal C}\gamma ^a$:
\begin{equation}
  {\cal C}^T= \sigma ({\cal C})\,{\cal C}\,,\qquad \left( {\cal C}\gamma ^a\right)
  ^T=\tau ({\cal C})\,{\cal C}\gamma ^a\,.
 \label{defsigtau}
\end{equation}
When we can define chiral spinors, called semi-spinors here, the
invariant $\iota $ indicates whether the charge conjugation matrix maps
between spinors of equal chirality $\iota(\beta )=1$ or different
chirality $\iota(\beta ) =-1$.

For \textbf{complex gamma matrices}, there are many references, and one
can compare e.g.\ with~\cite{VP}. The invariants $\sigma $ and $\tau $
and $\iota $ are related to the two numbers $\epsilon $ and $\eta $
of~\cite{VP} as
\begin{equation}
  \sigma ({\cal C})=-\epsilon \,,\qquad \tau ({\cal C} )=-\eta \,.
 \label{betatau}
\end{equation}
The main results depend on the dimension $n$. For \textbf{$n$ odd}, there
is one charge conjugation matrix (i.e.\ 1 bilinear form ${\cal C} $) and
\begin{eqnarray}
  n=1\mod 8 &:& \qquad \sigma ({\cal C} )=1\,,\qquad \tau ({\cal C} )=1\,,\nonumber\\
  n=3\mod 8 &:& \qquad \sigma ({\cal C} )=-1\,,\qquad\!\!\!\!\! \tau ({\cal C} )=-1\,,\nonumber\\
  n=5\mod 8 &:& \qquad \sigma ({\cal C} )=-1\,,\qquad\!\!\!\!\! \tau ({\cal C} )=1\,,\nonumber\\
  n=7\mod 8 &:& \qquad \sigma ({\cal C} )=1\,,\qquad \tau ({\cal C} )=-1\,.
 \label{stnodd}
\end{eqnarray}
For \textbf{even $n$} we can define a charge conjugation matrix for
either sign of $\tau $. We define
\begin{equation}
  \gamma_*\equiv (-\rmi)^{n/2+p}\gamma_1\ldots \gamma_n\,,\qquad
  \gamma_*\gamma_*=1\,.
 \label{defgamma*}
\end{equation}
Now, if ${\cal C}$ is a good charge conjugation matrix, then ${\cal
C}'={\cal C}\gamma_*$ is a charge conjugation matrix as well, with $\tau
({\cal C}')=-\tau ({\cal C})$. The value of $\sigma $ is
\begin{eqnarray}
&&  n=0\mod 8\ :\ \sigma ({\cal C} )=1\,,\qquad
  n=2\mod 8\ :\ \sigma ({\cal C} )=\tau ({\cal C} )\,,\nonumber\\
&&  n=4\mod 8\ :\ \sigma ({\cal C} )=-1\,,\qquad\!\!\!\!\!
  n=6\mod 8\ :\ \sigma ({\cal C} )=-\tau ({\cal C} )\,.
 \label{sigmaneven}
\end{eqnarray}
Using $(1\pm \gamma _*)/2$, we can define chiral spinors in this case,
and we find
\begin{equation}
 \iota ({\cal C} )=\stackrel{\leftarrow}{n}\equiv (-1)^{n(n-1)/2}\,.
 \label{iotacomplex}
\end{equation}
Here, $\stackrel{\leftarrow}{n}$ is the sign change on reversing $n$
indices of an antisymmetric tensor.

In this paper we more often make use of \textbf{real Clifford algebras}.
Explicit results on real Clifford algebras can be found in~\cite{O}. Here
we give some key results. Only in the cases $s=0, 6,7$, can the matrices
of the complex Clifford algebra (of dimension $2^{[n/2]}$, where the
Gauss bracket $[x]$ denotes the integer part of $x$) be chosen
to be real. This is called the \textit{normal type}.

In the other cases, we can get real matrices of dimension twice that of
the complex Clifford algebra. Many representations contain only pure real
or pure imaginary gamma matrices. A simple way to obtain real matrices of
double dimension is to use the matrices
\begin{equation}
  \Gamma ^a=\gamma^a\otimes \unity _2\quad \mbox{if }\gamma ^a\mbox{ is
  real}\,,\qquad
\Gamma ^a=\gamma^a\otimes \sigma_2\quad \mbox{if }\gamma ^a\mbox{ is
  imaginary}\,.
 \label{Gammareal}
\end{equation}
In the cases $s=1,5$, called the \textit{almost complex type}, there is
one matrix that commutes with all gamma matrices. It is
\begin{equation}
  J\equiv  \Gamma ^1\ldots \Gamma^n\,,\qquad J^2=-\unity \,.
 \label{defJ}
\end{equation}
Note that in the complex case the product of all the $\gamma^a$'s is
proportional to the identity for odd dimensions, but this is not so for
these larger and real $\Gamma^a$'s. (For $s=3,7$ the product of all real
gamma matrices is also $\pm \unity $). In this case there is always a
charge conjugation matrix $C$ with\footnote{In this explanation of
properties of real Clifford algebras, we will always denote by $C$ a
specific choice of charge conjugation matrix for the real Clifford
algebra, and by ${\cal C}$ the one for the complex gamma matrices. In
general we use ${\cal C}$ for any choice of charge conjugation
matrix.\label{fn:calCC}}
\begin{equation}
  \sigma (C)=\sigma ({\cal C})\,, \qquad \tau (C)=-1\,,
 \label{Cst15}
\end{equation}
where ${\cal C}$ indicates the charge conjugation matrix for the complex
case. There is also a matrix $D$ that satisfies the properties
\begin{equation}
  D\Gamma ^a+\Gamma ^aD=0\,,\qquad D^T=CDC^{-1}\,,\qquad \begin{array}{ll}
    D^2=\unity \quad & \mbox{if   }s=1\,, \\
    D^2=-\unity\quad  & \mbox{if }s=5 \,.
  \end{array}
 \label{propD}
\end{equation}

In the remaining cases, $s=2,3,4$, called the \textit{quaternionic type},
there are 3 matrices, which commute with all the $\Gamma^a$'s, denoted
$E_i$ for $i=1,2,3$. They satisfy
\begin{equation}
  [E_i,\Gamma ^a]=0\,,\qquad E_iE_j=-\delta _{ij}\unity +\varepsilon
  _{ijk}E_k\,,\qquad E_i^T=-CE_iC^{-1}\,,
 \label{Ei}
\end{equation}
where a charge conjugation matrix $C$ is used that satisfies
\begin{equation}
  \sigma (C)=-\sigma ({\cal C})\,, \qquad \tau (C)=\tau ({\cal C})\,.
 \label{Cst234}
\end{equation}

With these properties, we can obtain the following consequences for
bilinear forms.

\noindent \underline{$s=0,6$}. We have the normal type. The two charge
conjugation matrices of the complex Clifford algebra can be used
(possibly multiplied by $\rmi$ to make them real, but an overall factor
is not important), having opposite values of $\tau $. For $\sigma$ one
can use~(\ref{sigmaneven}). For $s=0$ there is no imaginary factor
in~(\ref{defgamma*}), and thus $\gamma _*$ is a real matrix that can be
used to define real chiral spinors (\textit{Majorana-Weyl spinors}). The
value of $\iota $ is then as in the complex case,
see~(\ref{iotacomplex}). For $s=6$ there is no projection possible in
this real case. The fact that the Clifford algebra is real reflects that
the irreducible spinors are \textit{Majorana spinors}.

\noindent \underline{$s=7$}. The real Clifford representation is also of
the normal type. With the odd dimension there is only one charge
conjugation matrix, and no chiral projection. The values of $\sigma $ and
$\tau $ are as in~(\ref{stnodd}). Again, the reality reflects the
property of \textit{Majorana spinors}.

\noindent \underline{$s=1,5$}. The real Clifford representation is of the
`almost complex type'. We have 4 choices for the charge conjugation
matrix: $C$, $CJ$, $CD$ and $CDJ$. We can derive from the given
properties that
\begin{eqnarray}
 &&\sigma ({\cal C})= \sigma (C)=-\stackrel{\leftarrow}{n}\sigma (CJ)=\sigma
  (CD)=\stackrel{\leftarrow}{n}\sigma (CDJ) \nonumber\\
  &&\tau (C)=\tau (CJ)=-1\,,\qquad \tau (CD)=\tau (CDJ)=1\,.
 \label{sigmataus15}
\end{eqnarray}
If $s=1$ then~(\ref{propD})  says that $\ft12(\unity \pm D)$ are good
projection operators, and can be used to define semispinors. These (real)
semispinors have the same dimension as the original complex ones and are
the \textit{Majorana spinors}. It is straightforward to check that
\begin{equation}
  \iota (C)=\iota (D)=1\,,\qquad \iota (CJ)=\iota (CDJ)=-1\,.
 \label{iotas1}
\end{equation}
If $s=5$ no such projection is possible. The size of the spinor
representation is doubled by the procedure~(\ref{Gammareal}), and this
reflects the fact that we have \textit{symplectic-Majorana} spinors.

\noindent \underline{$s=2,4$}. The real Clifford representation is of the
quaternionic type. With dimension even, we start from the two charge
conjugation matrices of the complex case. For each of them, we can
construct 3 extra ones by multiplying with the imaginary units $E_i$,
bringing the total to 8 invariant bilinear forms. From~(\ref{Ei})
and~(\ref{Cst234}) it follows that
\begin{eqnarray}
  &&\sigma ({\cal C})=-\sigma (C)= \sigma (CE_i) \nonumber\\
  &&\tau ({\cal C})=\tau (C)=\tau (CE_i)\,.
 \label{sts24}
\end{eqnarray}
The definition of chiral spinors as in the complex case is only possible
if $\gamma_*$ in~(\ref{defgamma*}) is real. Thus if $\ft12n +p$ is even,
i.e.\ $s=4$, the product of all the $\Gamma^a$'s is a good chiral
projection operator. The projected spinors are the components of
\textit{symplectic Majorana-Weyl spinors}. If $\gamma_*$ is imaginary,
i.e.\ $s=2$, the product of all the $\Gamma^a$'s squares to $-\unity $.
Using one of the complex structures, say $E_1$, then gives chiral
projections of the form $\ft12(\unity \pm \rmi\Gamma _*E_1)$. In this
case
\begin{equation}
  \iota (C)=\iota (CE_1)=-\iota({\cal C})\,,\qquad \iota (CE_2)=\iota
  (CE_3)=\iota ({\cal C})\,.
 \label{iotaquatE}
\end{equation}
The projected spinors are the components of \textit{Majorana spinors}.

\noindent \underline{$s=3$}. Here also, the real Clifford algebra is of
the quaternionic type, but since the dimension is odd, there is only one
charge conjugation matrix in the complex Clifford algebra. For
this,~(\ref{sts24}) applies. There is no chiral projection, and the
components correspond to \textit{symplectic Majorana spinors}.
\smallskip

The results can be seen in Table~\ref{tbl:beta} (though the names for the
different bilinear forms are unrelated to what has been explained here).
Table~\ref{tabA0} gives the number of solutions for charge conjugation
matrices that have $\beta =1$, $\beta =-1$.

\bigskip

The map $\Gamma ^k_\beta $ in the main text corresponds to the mapping
from two spinors $s$ and $t$ to the form with components $s^T{\cal
C}\Gamma _{a_1\ldots a_k}t$ (where ${\cal C}$ denotes now any choice as
explained in footnote~\ref{fn:calCC}), and the number $\sigma (\Gamma
^k_{{\cal C}})$ gives the symmetry of this bispinor (for commuting
spinors) under interchange of $s$ and $t$, while $\iota(\Gamma ^k_{{\cal
C}} )$ tells whether $s$ and $t$ have the same chirality. They are
related to $\sigma ({\cal C} )$, $\tau ({\cal C} )$ and $\iota ({\cal C}
)$ by~(\ref{sigmak}) and~(\ref{iotak}):
\begin{equation}
  \sigma (\Gamma ^k_{{\cal C}})=
  \sigma ({\cal C} )\tau ^k({\cal C} )\stackrel{\leftarrow}{k}\,,\qquad
  \iota (\Gamma ^k_{{\cal C}} )=(-)^k\iota ({\cal C} )\,.
 \label{sigiotak}
\end{equation}
For real Clifford algebras they are given explicitly in
Tables~\ref{tbl:sik1}--~\ref{tbl:sik3}.

\subsection{Summary of the results for the algebras}
This paper treats algebras that consist of an even sector $\gg_0
=\so(p,q) + W_0$, and an odd sector $\gg_1=W_1$ consisting of a
representation of $\so(p,q)$. The group $\so(p,q)$ is denoted as
$\so(V)$, and $V$ denotes its vector representation. We consider either
the usual case where the odd generators are fermionic ($\epsilon =1$, and
we have a superalgebra), or they can be bosonic ($\epsilon =-1$, and we
have a `$\bZ_2$-graded Lie algebra'). We will use the word `commutator'
in all cases, though this is obviously an anticommutator for $[W_1,W_1]$
in the superalgebra case. These algebras are called \textbf{$\epsilon
$-extensions of $\so(V)$.} We use the following terminologies for special
cases
\begin{description}
  \item[Poincar{\'e} superalgebras or Lie algebras:] $W_0$ are the
  translations in $n$ dimensions ($n=p+q$), which are denoted by $V$
   [and thus $\gg_0=\so(V)+V$] and $W_1$ is a spinorial representation.
  \item[Algebra of translational type:] all generators in $[W_1,W_1]$
  belong to $W_0$:
\begin{equation}
  [W_1,W_1]\subset W_0\,,\qquad [W_1,W_0]=0\,,\qquad [W_0,W_0]=0\,.
 \label{AlgTranslType}
\end{equation}
This part $W_0+W_1$ is called the `algebra of generalized translations'.
  \item[Transalgebra:] algebra of translational type where
  all the generators of $W_0$ appear in $[\gg_1,\gg_1]$, i.e.
\begin{equation}
  [W_1,W_1]=W_0\,.
 \label{Transalgebra}
\end{equation}
  \item [$\epsilon $-extended polyvector Poincar{\'e} algebras:] Algebra of
  translational type where $W_1$ is a (possibly reducible) spinorial
  representation (includes chiral and extended supersymmetry).
\end{description}
  There are 2 extreme cases: one in which the full $\gg$ is semisimple,
  which is the case of the Nahm superalgebras,
  and the \textbf{algebras of semi-direct type}, where $\so(p,q)$ is its largest
  semisimple subalgebra.

Apart from degenerate cases where $n\leq 2$, any transalgebra is of
semi-direct type.

Transalgebras are \textit{minimal} cases of algebras of translational
type in the sense that there are no proper subalgebras, see
definition~\ref{superextDef}. In fact, any algebra of translational type
can be written as $\gg=\gg'+\ga$, where $\ga\subset W_0$ is an $\so(p,q)$
representation, which is irrelevant in the sense that all its generators
commute with all of $W_1$ and $W_0$ and do not appear in $[W_1,W_1]$. The
algebra $\gg'$ is a transalgebra.

For any choice of $W_1$ there is a unique transalgebra where $W_0$ has
all the $\so(p,q)$ representations that appear in the (anti)symmetric
product of $W_1$ with itself. The (anti)commutators of $W_1$ are then
\begin{equation}
  [W_1,W_1]=\sum_{\mbox{all }r}  W_0^{(r)}\,,
 \label{W1W1basicTransalg}
\end{equation}
Here $r$ labels all representations that appear in the symmetric product
for $\epsilon =1$, i.e.\ superalgebras, and in the antisymmetric product
for $\epsilon =-1$, i.e. for Lie algebras.

Any other transalgebra can be obtained by removing an arbitrary number of
terms in (\ref{W1W1basicTransalg}). We can consider these to be
contractions of this basic transalgebra, where the representations to be
removed are multiplied by some parameter $t$ and the limit $t\rightarrow
0$ is taken.

Any $\epsilon $-extension of semi-direct type with $W_1$ irreducible and
of dimension at least 3, is of the following form
\begin{eqnarray}
 &&W_0  =  A + K\,,\qquad [K,W_1]=0\,,\qquad [A,W_1]= \rho W_1\,, \nonumber\\
 &&[\so(V),A]=0 \,,\qquad [A,A]\subset K\,,\qquad  [W_1,W_1]\subset K\,.
 \label{semidirect}
\end{eqnarray}
where $\rho =0$ or $\bR{\cdot}{\rm Id}$ or $\bC{\cdot}{\rm Id}$. This is thus a
transalgebra iff $A=0$. Also, if the algebra is minimal, then
$[W_1,W_1]=W_0$ and it is a transalgebra.

Also, if $W_0$ and $W_1$ are irreducible $\so(p,q)$ representations, then
either the algebra is of translational type, i.e.\ $[W_0,W]=0$, or  $W_0$
is an abelian generator and $[W_0,W_1]=a\,W_1$, where $a$ is a number.

We now restrict ourselves to transalgebras where $W_1=S$, the irreducible
spinor representation of $\so(p,q)$. Then the representations that appear
in the right-hand side of~(\ref{W1W1basicTransalg}) are either $k$-forms
or, in the case that $s=p-q$ is divisible by 4,  also (anti-)selfdual
$(n/2)$-forms. Thus, the unique maximal transalgebra has
(anti)commutators
\begin{equation}
  \left[S_\alpha,S_\beta \right] =
   \sum_{k} ({\cal C}\Gamma^{a_1\ldots a_k})_{\alpha \beta } W_0^{k}{}_{\,a_1\ldots a_k}\,,
 \label{SStransalgebra}
\end{equation}
where $\alpha ,\beta $ denote spinor indices. The classification of
transalgebras with $W_1=S$ reduces to the description of all the charge
conjugation matrices ${\cal C}$ and the specification of the range of the
summation over $k$. The relevant issue is the symmetry for a particular
$k$, i.e.\ the $\sigma (\Gamma ^k_{{\cal C}})$ of the previous
subsection. When there are chiral spinors involved, the chirality should
be respected, which is related to $\iota(\Gamma ^k_{{\cal C}})$.

\begin{table}[p]
  \begin{center}
  \begin{tabular}{|l|l|l|}
\hline
   & Superalgebra: $\sigma =+1$ & Lie algebra: $\sigma =-1$ \\
\hline
 $n=2m+1$ & $k=m-4i$ & $k=m-1-4i$ \\
   & $k=m-3-4i$ & $k=m-2-4i $\\
\hline
 $n=2m $&   &   \\{}
 $[S_\pm ,S_\pm ]$ & $k=m-4-4i$ & $k=m-2-4i$ \\{}
   & $k=m $ \hfill (anti)selfdual &   \\{}
 $[S_+,S_-] $& $k=m-1-2i$ &$ k=m-1-2i $  \\\hline
\end{tabular}
\end{center}
\caption{\it The values of $k$ in~(\ref{SStransalgebra}) for the case of
  complex spinors. $n$ is the dimension of the vector space. $i$ can be $0,1,\ldots $
  limited by the fact that obviously $k\geq 0$. For the even-dimensional case,
  we split the (anti)commutator between spinors of different and equal chirality.
  For equal chirality, the $k=m$ generator is either selfdual or antiselfdual. }
  \label{tbl:kcomplex}
  \end{table}

\begin{table}[p]
\begin{center}
 $ \begin{array}{|l|l|l|}
\hline
   & \mbox{Superalgebra}: \sigma =+1 & \mbox{Lie algebra}: \sigma =-1 \\
\hline\hline
 \mathbf{n=2m+1} &&\\
 s=1,\,7\ (\mbox{M}) & k=m-4i & k=m-1-4i \\
   & k=m-3-4i & k=m-2-4i \\
\hline
 s=3,\,5\ (\mbox{SM})
  & k=m-4i\hfill\mbox{triplet} & k=m-4i\hfill \mbox{singlet}\\
 & k=m-3-4i\hfill \mbox{triplet} & k=m-3-4i\hfill \mbox{singlet}\\
& k=m-1-4i \hfill \mbox{singlet} & k=m-1-4i\hfill  \mbox{triplet}\\
& k=m-2-4i\hfill \mbox{singlet} & k=m-2-4i\hfill \mbox{triplet}\\
 \hline \hline
 \mathbf{n=2m }&   &   \\{}
 s=0\ (\mbox{MW}) & &\\{}
 [S_\pm ,S_\pm ] & k=m-4-4i & k=m-2-4i \\{}
   & k=m\hfill \mbox{ (anti)selfdual} &   \\{}
 [S_+,S_-] & k=m-1-2i & k=m-1-2i  \\ \hline
s=2,\,6\ (\mbox{M}) & k=m-4i,\,m+4+4i & k=m-1-4i,\,m+3+4i\\
                 &k=m-3-4i,\,m+1+4i & k=m-2-4i,\,m+2+4i\\ \hline
s=4\ (\mbox{SMW})&&\\{}
 [S_\pm ,S_\pm ]   & k=m-4-4i\hfill\mbox{triplet} & k=m-2-4i\hfill\mbox{triplet}\\
 & k=m\quad\hfill\mbox{(anti)selfdual triplet}& k=m\quad\hfill\mbox{(anti)selfdual singlet}\\
& k=m-2-4i\hfill\mbox{singlet} & k=m-4-4i\hfill \mbox{singlet}\\
{} [S_+,S_-] & k=m-1-2i \hfill 2\times 2
 & k=m-1-2i \hfill 2\times 2  \\
   \hline\hline
\end{array}$
\end{center}
\caption{\it The values of $k$ in~(\ref{SStransalgebra}) for the case of
  real spinors. $n$ is the dimension of the vector space. $i$ can be $0,1,\ldots$
  limited by the fact that obviously $k\geq 0$ (and $k\leq n$ for $s=2,6$).
  In cases where there are real Weyl spinors,
  we split the (anti)commutator between spinors of different and equal
  chirality, and the $k=m$ generator is either selfdual or antiselfdual. When there are
  symplectic spinors, the right-hand side of~(\ref{SStransalgebra}) contains for some
  $k$'s triplets of the automorphism group $\su(2)$, and singlets for other $k$'s. The types of real
   spinors, Majorana (M), symplectic-Majorana (SM), or symplectic Majorana-Weyl (SMW)
   are indicated.}
  \label{tbl:kreal}

\end{table}

First, in Section~\ref{complex}, the complex case is discussed. That
means that there are no reality conditions on bosonic or fermionic
generators. When the dimension is odd, the result is given in
Theorem~\ref{th_odd}. There is only one charge conjugation matrix, and
the result can be understood from~(\ref{sigiotak}) and~(\ref{stnodd}).
For even dimensions the result is given in Theorem~\ref{th_even}. This
depends mainly on~(\ref{sigmaneven}) and~(\ref{iotacomplex}). Here the
spinors can be split into chiral spinors, and we can separately consider
the commutators between spinor generators of the same and of opposite
chirality. The result for allowed values of $k$ in~(\ref{SStransalgebra})
can be found also in Table~\ref{tbl:kcomplex}.

As an example we may check that in 11 dimensions we can indeed have $P$,
$Z^{ab}$ and $Z^{a_1\ldots a_5}$ generators in $W_0$, as is the case of
the M-algebra, and the classification implies that we can consistently
put any one of these to zero.

For the case of real generators, it is important to note that
(anti)selfdual tensors in even dimensions are only consistent for $s/2$
even. We now discuss the algebras according to the 8 different values of
$s$. The results are shown in Table~\ref{tbl:kreal}.

\noindent \underline{$s=0$} (Majorana-Weyl spinors). There are chiral
spinors and we can split the commutators. The $k$ values that appear in
Tables~\ref{tbl:beta}--~\ref{tbl:sik3} with $\iota =1$ can appear in
commutators of equal chirality. The value of $\sigma $ indicates whether
they appear in superalgebras ($\sigma =1$) or in Lie algebras ($\sigma
=-1$). Those with $\iota= -1$ appear in the same way in commutators of
different chirality. The (anti)selfdual tensors appear in the commutators
between spinors of the same chirality.

\noindent \underline{$s=1$} (Majorana spinors). The two projections to
semispinors mentioned above~(\ref{iotas1}), lead to equivalent spinors.
We thus consider only the commutator between these irreducible spinors
(including the others is contained in the `extended algebras' discussed
below). In Tables~\ref{tbl:beta}--~\ref{tbl:sik3} we thus consider the
$\iota =1$ cases. We can check that $\iota =-1$ always allows both
$\sigma=1$ and $\sigma =-1$ as this concerns commutators between
unrelated but equivalent spinors.

\noindent \underline{$s=2$} (Majorana spinors). The two projections to
semispinors lead to equivalent spinors. We thus consider only the
commutator between these irreducible spinors. Note that in the table we
indicate here also forms with $k>m$. These are dual to $k<m$ forms, and
this duality has been used in the formulation of the $s=2$ part of
Theorem~\ref{thm:6}. The formulation here shows the gamma matrices
completely, e.g. the appearance of $\Gamma _{abc}= \varepsilon
_{abcd}\gamma _5\gamma ^d$ in 4 dimensions.

\noindent \underline{$s=3$, $s=5$} (Symplectic-Majorana spinors). The
symplectic spinors are in a doublet of $\su(2)$. According to the value
of $\sigma $ for a particular $k$ we find either a triplet or a singlet
of generators in the superalgebra or in the Lie algebra.

\noindent \underline{$s=4$} (Symplectic Majorana-Weyl spinors). In the
commutators between generators of equal chirality (which are again
doublets of $\su(2)$), we find either triplets (symmetric) or singlet
(antisymmetric) generators. For commutators between generators of
different chirality no symmetry or antisymmetry can be defined, and the
generators allowed by the chirality ($\iota =-1$) appear in the
superalgebra as well as in the Lie algebra.

\noindent \underline{$s=6$} (Majorana spinors). This case is
straightforward from the tables and the spinors are just real and not
projected. Remark that the result is then the same as for the projected
ones of $s=2$. The same remark about showing tensors with $k>m$ holds
here too. These are dualized in the formulation in Corollary~\ref{cor4}.

\noindent \underline{$s=7$} (Majorana spinors). Here also, the tables
straightforwardly lead to the same result as for the projected spinors of
$s=1$.
 \medskip

We remark that the result is the same for $s$ and for $-s$, which shows
that the conventional choices discussed at the beginning of
Section~\ref{app:ClComReal} do not influence the final algebras.
\bigskip

Finally, in Section \ref{N-extention}, results are obtained for
\textbf{$\boldsymbol{N}$-extended polyvector Poincar{\'e} algebras}. This
means that $W_1$ consists of $N$ copies of the irreducible spinor $S$. In
cases where there are two inequivalent copies (complex even dimensional,
or real with $s=0$ or $s=4$) we have
\textbf{$\boldsymbol{(N_+,N_-})\boldsymbol{(N_+,N_-})$-extended
polyvector Poincar{\'e} algebras}.

The results are straightforward from the above tables and this shows why
it has been useful to include the Lie algebra case. The generators in
$W_1$ are in an $N$-representation of the automorphism algebra that acts
on the copies of $S$.

\textit{For the complex odd-dimensional case and real $s=1,2,6,7$
(Majorana)}:
 We just have to split the $N\times N$
representations into the symmetric and antisymmetric ones.
\begin{eqnarray}
  \mbox{for superalgebras:}&\qquad& \fr{N(N+1)}{2}
  \mbox{ copies of the $\sigma =1$ generators}
  \nonumber\\
&&  +\fr{N(N-1)}{2}\mbox{ copies of the $\sigma =-1$ generators}\nonumber\\
\mbox{for Lie algebras:~~}&\qquad& \fr{N(N-1)}{2}\mbox{ copies of the
$\sigma =1$ generators}\nonumber\\
&&  +\fr{N(N+1)}{2}\mbox{ copies of the $\sigma =-1$ generators}
\label{Nextended}
\end{eqnarray}

\textit{For the complex even-dimensional case and real $s=0$ (Weyl)}: We
have $(N_+,N_-)$ algebras. We use the above rule separately for the
commutators between the $N_+$ chiral generators and between the $N_-$
antichiral ones. Furthermore there are $N_+N_-$ copies of the generators
that appear in $[S_+,S_-]$ in Tables~\ref{tbl:kcomplex}
and~\ref{tbl:kreal}. As an example, the $(2,1)$ superalgebra in
8-dimensional (4,4) space contains: three selfdual 4-forms, and one
antiselfdual 4-form, four 0-forms (three in $[2S_+,2S_+]$ and one in
$[S_-,S_-]$, one 2-form (in $[2S_+,2S_+]$) and two 3-forms and 1-forms in
$[2S_+,S_-]$.

\textit{For the symplectic real case $s=3,5$}: The automorphism algebra
is already $\gsp(2)=\su(2)$ for the simple algebras discussed above. For
the extended algebras it is $\gsp(N)$ where $N$ is even. The simple case
is thus similar to~(\ref{Nextended}) with $N=2$, and the `triplet' and
`singlet' indications in Table~\ref{tbl:kreal} reflect this. Therefore
for higher $N$ (always even) we replace in Table~\ref{tbl:kreal} the
`triplet' by $N(N+1)/2$ and the `singlet' by $N(N-1)/2$.

\textit{For the symplectic Majorana-Weyl case $s=4$}: We merely need to
combine the remarks above for the symplectic case and the Weyl case.
Extended algebras are of the form $(N_+,N_-)$ where both numbers are
even. The `triplet' indication in Table~\ref{tbl:kreal} is replaced by
$N_+(N_++1)/2$ and $N_-(N_-+1)/2$ and `singlet' is replaced by
$N_+(N_+-1)/2$ and $N_-(N_--1)/2$. The mixed commutators are multiplied
by $N_+N_-$.

\newpage
%%%%%%%%%%%%%%%%%%%%%%%%%%%%%%%%%%%%%%%%%%%%%%%%%%%%%%%%

%%%%%%%%%%%%%%%%%%%%%%%%%%%%%%%%%%%%%%%%%%%%%%%%%%%%%%%
\end{document}